\documentclass{aa}                       
\usepackage{ amsmath }
\usepackage{ amssymb }
\usepackage{graphicx}
\usepackage{aas_macros}
\usepackage{subcaption}
\usepackage{morefloats}
\usepackage[varg]{txfonts}
\usepackage{fixltx2e}
\usepackage[3D]{movie15}
\usepackage{hyperref}
\usepackage{epsfig}

\pdfoutput=1

\bibpunct{(}{)}{;}{a}{}{,}

\begin{document}
\title{Mid-J CO shock tracing observations of infrared dark clouds II}
\subtitle{Low-J CO constraints on excitation, depletion, and kinematics \thanks{Based on observations carried out with the JCMT and IRAM 30m Telescopes. IRAM is supported by INSU/CNRS (France), MPG (Germany) and IGN (Spain).}}
\titlerunning{Mid-J CO shock tracing observations of IRDCs II}

\author{A. Pon\inst{\ref{MPE},\ref{UWO}}
\and D. Johnstone\inst{\ref{JAC},\ref{HIA},\ref{UVic}}
\and P. Caselli\inst{\ref{MPE}} 
\and F. Fontani\inst{\ref{INAF}}
\and A. Palau\inst{\ref{CRyA}}
\and M. J. Butler\inst{\ref{MPIA}}
\and M. Kaufman\inst{\ref{SanJose},\ref{Ames}}
\and I. Jim\'{e}nez-Serra\inst{\ref{UCL}} 
\and J. C. Tan\inst{\ref{Florida}}}

\institute{Max-Planck-Institut f\"{u}r extraterrestrische Physik, Giessenbachstrasse 1, D-85748 Garching, Germany\label{MPE}
\and Department of Physics and Astronomy, The University of Western Ontario, London, Canada, N6A 3K7 \email{apon@uwo.ca}\label{UWO}
\and Joint Astronomy Centre, 660 North A'ohoku Place, University Park, Hilo, HI 96720, USA\label{JAC}
\and NRC-Herzberg Institute of Astrophysics, 5071 West Saanich Road, Victoria, BC V9E 2E7, Canada\label{HIA}
\and Department of Physics and Astronomy, University of Victoria, PO Box 3055 STN CSC, Victoria, BC V8W 3P6, Canada\label{UVic}
\and INAF- Osservatorio Astrofisico di Arcetri, Largo E. Fermi 5, Firenze I-50125, Italy\label{INAF}
\and Instituto de Radioastronom\'ia y Astrof\'isica, Universidad Nacional Aut\'onoma de M\'exico, P.O. Box 3-72, 58090 Morelia, Michoac\'an, M\'exico\label{CRyA}
\and Max Planck Institute for Astronomy, K\"{o}nigstuhl 17, 69117 Heidelberg, Germany\label{MPIA}
\and Department of Physics and Astronomy, San Jose State University, One Washington Square, San Jose, CA 95192-0106\label{SanJose}
\and Space Science and Astrobiology Division, MS 245-3, NASA Ames Research Center, Moffett Field, CA 94035, USA\label{Ames}
\and Department of Physics and Astronomy, Unviersity College London, 132 Hampstead Road, London NW1 2PS, UK\label{UCL}
\and Departments of Astronomy \& Physics, University of Florida, Gainesville, FL 32611, USA\label{Florida}}

\abstract{Infrared dark clouds are kinematically complex molecular structures in the interstellar medium that can host sites of massive star formation. We present 4 square arcminute maps of the $^{12}$CO, $^{13}$CO, and C$^{18}$O J = 3 to 2 lines from selected locations within the C and F (G028.37+00.07 and G034.43+00.24) infrared dark clouds (IRDCs), as well as single pointing observations of the $^{13}$CO and C$^{18}$O J = 2 to 1 lines towards three cores within these clouds. We derive CO gas temperatures throughout the maps and find that CO is significantly frozen out within these IRDCs. We find that the CO depletion tends to be the highest near column density peaks, with maximum depletion factors between 5 and 9 in IRDC F and between 16 and 31 in IRDC  C. We also detect multiple velocity components and complex kinematic structure in both IRDCs. Therefore, the kinematics of IRDCs seem to point to dynamically evolving structures yielding dense cores with considerable depletion factors.}


\maketitle


\section{INTRODUCTION}
\label{introduction}


Infrared dark clouds (IRDCs) are dense, molecular structures that appear dark against the bright mid-infrared ($\sim 8 \mu$m) Galactic background \citep{Perault96, Egan98}. IRDCs are often filamentary and some are associated with high mass star formation (e.g., \citealt{Rathborne06, Busquet13}). Previous observations of IRDCs show that they are kinematically complex, with numerous filamentary substructures (e.g., \citealt{Molinari10, Henshaw13, JimenezSerra14, Henshaw14}). IRDCs can contain multiple separate velocity components that are not obvious simply from spatial information (e.g., \citealt{Henshaw13, Henshaw14}). These large-scale kinematic signatures are key pieces of evidence in deciphering the importance of global, dynamical motions to the process of star formation, potentially revealing the influence of colliding flows (e.g., \citealt{VazquezSemadeni09, Dobbs14}), cloud-cloud collisions (e.g., \citealt{Scoville86, Tan00, Wu15VanLoo}), and global gravitational collapse (e.g., \citealt{BallesterosParedes11Hartmann}) in producing not only prestellar cores, but the larger molecular clouds themselves.

At the high densities ($n_{\text{H}}$ $>$ 10$^5$ cm$^{-3}$) and low temperatures (T $\lesssim$ 20K) prevalent in the centers of prestellar cores, CO freezes out of the gas phase onto dust grains (e.g., \citealt{Caselli99}). The CO depletion factor, $f_\text{D}$, is defined as the ratio of the column density of gas estimated via a non-CO based method, such as from dust continuum emission or extinction, to the total column density of gas derived from the integrated intensities of CO rotational transitions, under the assumption of a typical CO gas phase abundance of 1 to $2 \times 10^{-4}$ relative to H$_2$. This is equivalent to the ratio between the expected gas phase CO abundance in the absence of grain surface processes to the observed gas phase CO abundance. Towards dense cores in low mass star-forming regions, depletion factors of 5 to 15 are commonly found (e.g., \citealt{Caselli99, Crapsi05}), but the observational evidence for prevalent, significant CO depletion is less clear in IRDCs. While some observations find CO depletion factors similar to that in low mass cores, with depletion factors of the order of 5 to 10 (e.g., \citealt{Hernandez11Caselli, Hernandez12, JimenezSerra14}), other studies of IRDCs reveal depletion factors ranging from one, indicating no significant CO freeze out \citep{Miettinen11, Hernandez11Tan}, to 80 \citep{Fontani12}. 

The dynamic environment of IRDCs can lead to significant thermal processing of gas. Low velocity shocks can significantly heat small regions of gas (e.g., \citealt{Pon12Kaufman}) and widespread SiO emission, typically considered to be a tracer of shocks, has been detected in multiple IRDCs \citep{JimenezSerra10, NguyenLuong13}. \citet[hereafter Paper I]{Pon15Caselli}, used the {\it Herschel Space Observatory} to observe four reasonably quiescent clumps within IRDCs in an attempt to detect enhanced mid-J CO emission indicative of a warm gas component created by shocks within the IRDCs. Enhanced emission was detected in the CO J = 8 $\rightarrow$ 7 and 9 $\rightarrow$ 8 lines for three clumps, the C1, F1, and F2 clumps of \citet{Butler09}. The characterization of this mid-J CO emitting gas was limited, however, by a lack of lower-J CO observations constraining the properties of the cooler gas within the IRDCs that accounts for the majority of the mass. In this paper, we present low-J CO observations of the C1, F1, and F2 clumps in order to constrain the bulk properties of the gas in and around these clumps. In Paper III of this series (Pon et al.\ in preperation), we will combine the low-J and mid-J observations to evaluate the properties of the warm gas component in these clumps.

The C1, F1, and F2 clumps are reasonably well-studied, dense clumps embedded within IRDCs (e.g., \citealt{Rathborne06, Carey09,Fontani11,Butler12, Tan13}; Paper I). The C1 clump is within IRDC C (G028.37+00.07) and the F1 and F2 clumps are within IRDC F (G034.43+00.24) of the \citet{Butler09} sample. IRDC C is also known as the Dragon Nebula \citep{Wang15Thesis}. The C1 clump contains two massive prestellar core candidates, the C1-N and C1-S cores, while the F1 and F2 clumps contain the F1 and F2 prestellar cores, respectively. As in Paper I, we refer to structures that should fragment into clusters of stars as clumps and structures that should give rise to individual stars or small groups as cores. As such, we interpret objects on the size scale of $\sim$ 1 pc as clumps and structures on the scale of $\sim$ 0.1 pc as cores. Information about the four cores is given in Table \ref{table:targets}.

\begin{table*}
\begin{minipage}{\textwidth}
\caption{Target Locations}
\begin{center}
\begin{tabular}{cccccccccc}
\hline
\hline
Core & Clump & Cloud & RA (J2000) & Dec(J2000) & $l$ & $b$ & $V_\text{core}$ & $V_\text{cloud}$ & $d$ \\
 & & & (h:m:s) & ($^\circ$:\arcmin:\arcsec) & ($^\circ$) & ($^\circ$) & (km s$^{-1}$) & (km s$^{-1}$) & (kpc) \\
(1) & (2) & (3) & (4) & (5) & (6) & (7) & (8) & (9) & (10) \\
\hline
C1-N & C1 (MM9) &  G028.37+00.07 (C) & 18:42:46.9 & -04:04:06 & 28.32503 & 0.06724 & 81.18 & 78.6 & 5.0 \\
C1-S & C1 (MM9) & G028.37+00.07 (C) & 18:42:46.5 & -04:04:16 & 28.32190 & 0.06745 & 79.40 & 78.6 & 5.0 \\
F1 & F1 (MM8) & G034.43+00.24 (F) & 18:53:16.5 & 01:26:09 & 34.41923 & 0.24598 & 56.12 & 57.1 & 3.7 \\
F2 & F2 & G034.43+00.24 (F) & 18:53:19.2 & 01:26:53 & 34.43521 & 0.24149 & 57.66 & 57.1 & 3.7 \\
\hline
\end{tabular}
\tablefoot{Column 1 gives the name of the core as denoted by \citet{Tan13}. Column 2 gives the name of the parent clump as denoted by \citet{Butler12} and, in parenthesis, the original name of the clump as assigned by \citet{Rathborne06}. The F2 clump was not given a designation by \citet{Rathborne06}. Column 3 gives the name of the IRDC in which the core is embedded. Columns 4 and 5 give the right ascension and declination of the core and Cols.~6 and 7 give the Galactic longitude and latitude of the core, based on the N$_2$D$^+$ detections of \citet{Tan13}. Column 8 gives the central velocity, with respect to the local standard of rest, of the core, as measured with the N$_2$D$^+$ J = $3 \rightarrow 2$ line observed by \citet{Tan13}. The local standard of rest velocity of the parent cloud, from \citet{Simon06}, is given in Col.~9. Column 10 gives the kinematic distance of the cloud from \citet{Simon06}. While \citet{Kurayama11} find a 1.56 kpc distance for IRDC F, containing F1 and F2, \citet{Foster12} find a distance consistent with the kinematic distance for IRDC F, based upon extinction measurements. We thus elect to use the kinematic distances from \citet{Simon06} for both IRDCs in this paper.}
\label{table:targets}
\end{center}
\end{minipage}
\end{table*}

\begin{figure*}
   \centering
   \begin{subfigure}[b]{0.5\textwidth}
      \centering
       \includegraphics[height=3.6in, angle=270]{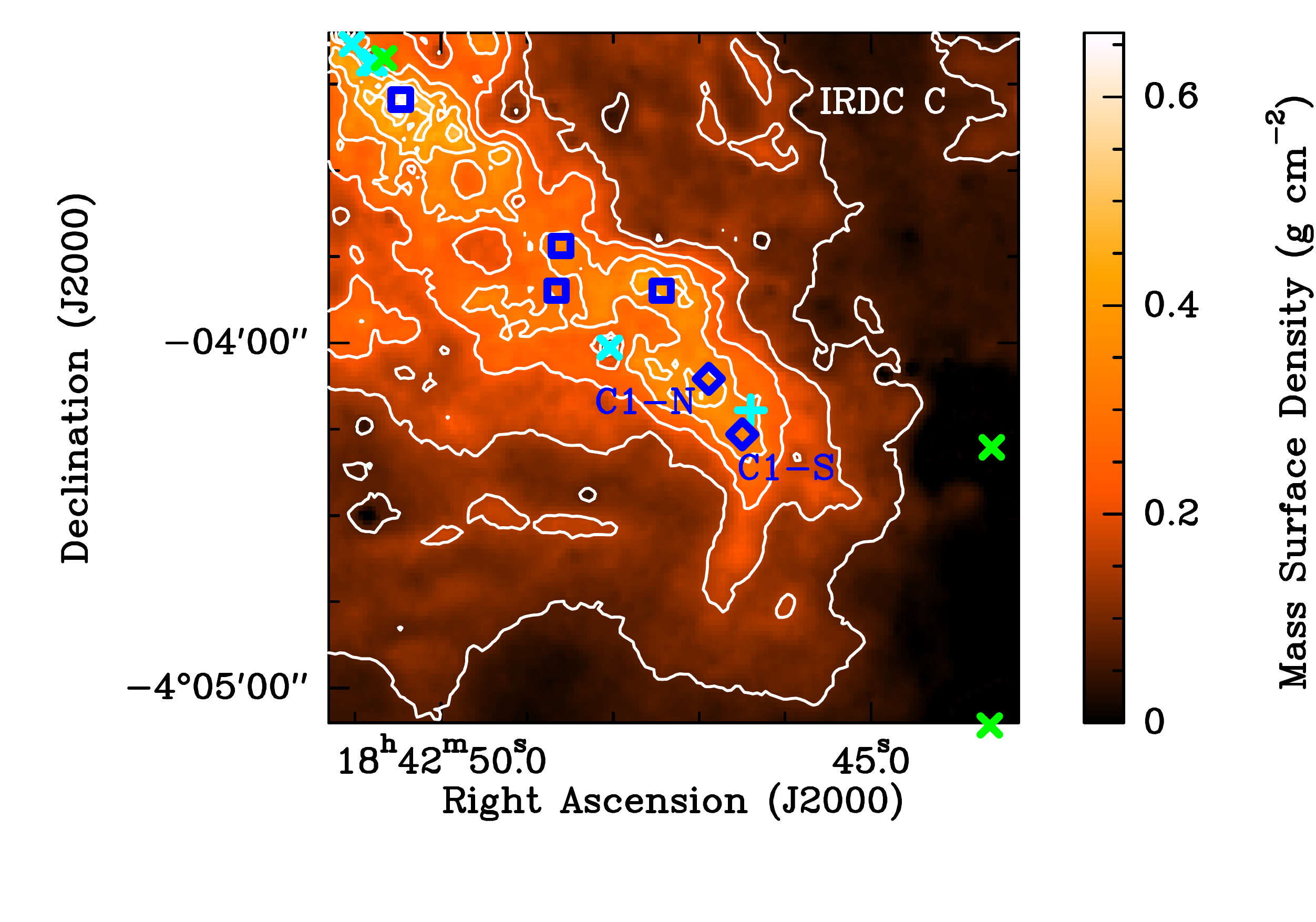}
   \end{subfigure}%
  \begin{subfigure}[b]{0.5\textwidth}
     \centering
     \includegraphics[height=3.6in, angle=270]{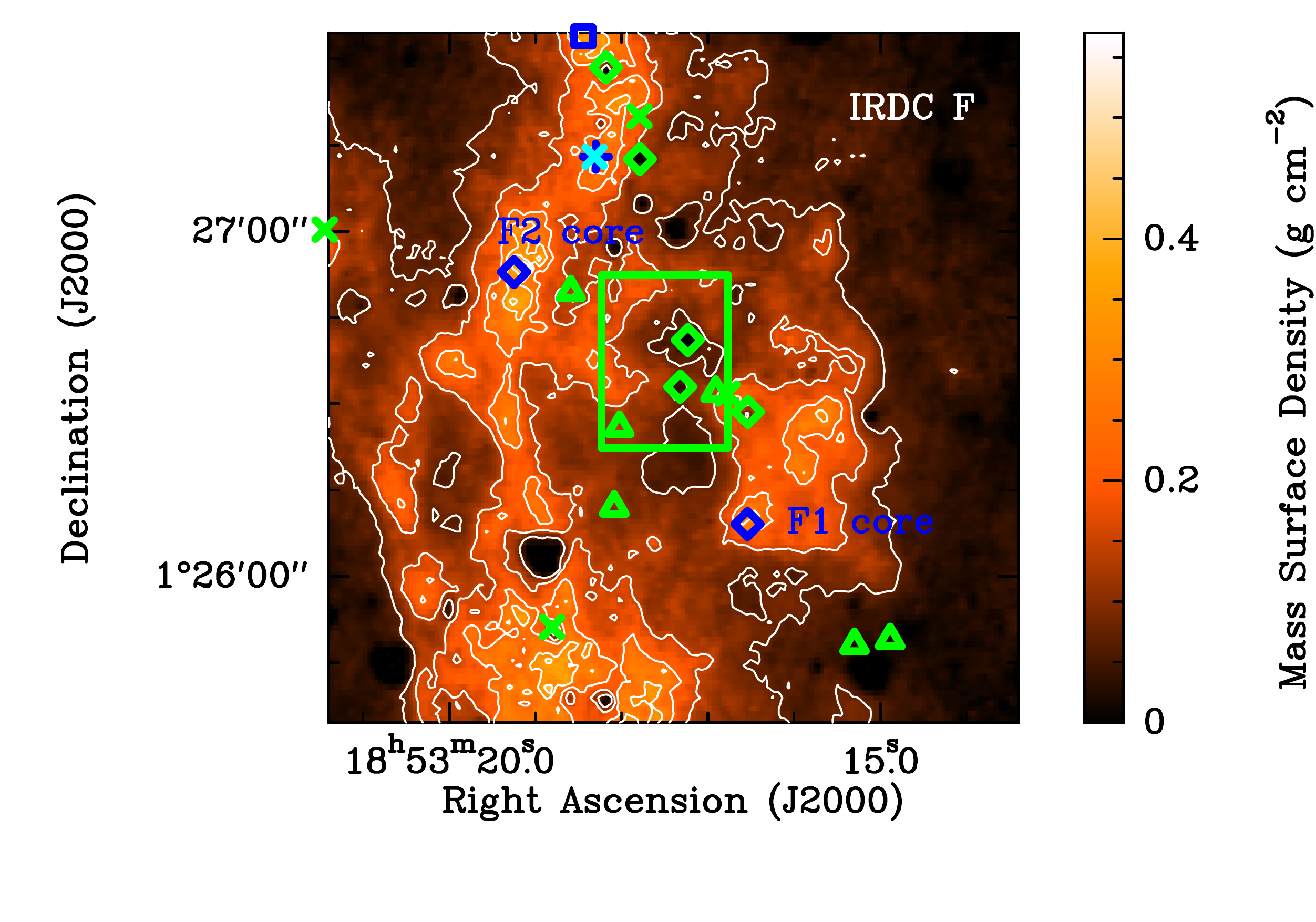}
   \end{subfigure}
   \caption{Mass surface density derived by \citet{Butler12} for the C1 clump (left) and F1 and F2 clumps (right). The contours start at 0.075 g cm$^{-2}$ ($A_\text{V}$ of 17 mag) and increase by increments of 0.075 g cm$^{-2}$, with the typical uncertainty being on the order of 30\% based upon uncertainties in the opacity per unit mass \citep{Butler09,Butler12}. The uncertainty based upon background fluctuations is of the order of 0.013 g cm$^{-2}$ \citep{Butler12}. The dark blue diamonds give the central locations of the C1-N, C1-S, F1, and F2 cores, as determined from the N$_2$D$^+$ ALMA observations of \citet{Tan13}. The dark blue squares give the locations of extinction cores identified by \citet{Butler14}. The central location of the \citet{Butler14} C1 clump is not shown in lieu of the \citet{Tan13} locations. The light blue X's indicate the positions of water masers \citep{Wang06, Chambers09, Wang12} while the light blue cross indicates the position of the water maser detected at 59 km s$^{-1}$ by \citet{Wang06} but not detected by the more sensitive survey of \citet{Chambers09}. The green X's indicate the positions of 70 $\mu$m sources from the HiGAL survey \citep{Molinari10}. The green diamonds indicate positions of objects that were well fit as young stellar objects (YSOs) by \citet{Shepherd07}. The green triangles are 24 $\mu$m sources that are likely YSOs, but which could also be asymptotic giant branch stars since they are only detected in two or less of the IRAC+2MASS bands  \citep{Shepherd07}. The green rectangle indicates the location of a near-infrared source overdensity (overdensity A), interpreted by \citet{Foster14} as an embedded low mass protostar population. Note that the right ascension for this overdensity given in Table 3 of \citet{Foster14} is incorrect (Foster, personal communication) and we use a right ascension of 18$^h$53$^m$17$\fs$5 (J2000), not 18$^h$53$^m$20$\fs$5 (J2000), for this overdensity. The regions shown are the regions mapped by the JCMT.}
   \label{fig:sigmaradec}
\end{figure*}

Figure \ref{fig:sigmaradec} shows the mass surface densities of the three clumps, as derived from mid-infrared extinction data by \citet{Butler12}. This figure also shows the locations of various embedded sources within the clouds. 

Because the F1, F2, C1-N, and C1-S cores are cool and dense, they should be prime candidates for objects exhibiting large CO depletions. Their parental clumps also present an ideal opportunity to examine the kinematic structure of gas surrounding nascent star-forming sites, before the surrounding gas is significantly impacted by feedback from newly formed protostars. 

In Sect.~\ref{JCMT reduction}, we present James Clerk Maxwell Telescope (JCMT) derived maps of the $^{12}$CO, $^{13}$CO, and C$^{18}$O J = 3 $\rightarrow$ 2 transitions towards the C1, F1, and F2 clumps. In Sect.~\ref{Fontani}, we then present IRAM 30m single pointing observations of the $^{13}$CO and C$^{18}$O J = 2 $\rightarrow$ 1 transitions towards the C1-N, F1, and F2 cores. We derive the excitation temperature of the J = 3 and J = 2 states of $^{12}$CO, based upon the $^{12}$CO J = 3 $\rightarrow$ 2 line, in Sect.~\ref{temperature} and compare the kinetic gas temperatures traced by the $^{13}$CO and $^{12}$CO lines. The CO depletion factor towards our observed areas is then derived and discussed in Sect.~\ref{depletion}. The kinematics of these IRDCs is then analyzed in Sect.~\ref{kinematics}. Finally, the primary conclusions are summarized in Sect.~\ref{conclusions}. The full spectral data sets are provided in Appendix \ref{appendix:jcmtfits}, available in the online version. 

\section{OBSERVATIONS}
\label{observations}

\subsection{JCMT Data and Reduction}
\label{JCMT reduction}

As part of the director's discretionary time program M13AD02 on the James Clerk Maxwell Telescope (JCMT), maps of the $^{12}$CO, $^{13}$CO, and C$^{18}$O J = 3 $\rightarrow$ 2 transition were created for the regions around the C1, F1, and F2 clumps. Due to the proximity of F1 and F2, both of these clumps were observed in a single map. For each IRDC, a 16 x 16 pixel map (112.5 arcsec per side, 7.5 arcsec spacing) was made using the 4x4 Jiggle-chop mode of the Heterodyne Array Receiver Programme (HARP) instrument \citep{Buckle09}. The Autocorrelation Spectrometer and Imaging System (ACSIS) was used as the backend \citep{Buckle09}. The $^{13}$CO and C$^{18}$O observations were made simultaneously with 61 kHz spectral resolution (roughly 0.05 km s$^{-1}$), while the $^{12}$CO data were taken with 31 kHz spectral resolution (roughly 0.03 km s$^{-1}$). The JCMT half power beam width (HPBW) was approximately 15 arcsec for these observations. To convert from antenna temperatures to main beam temperatures, a conversion factor of 0.61 is adopted for all observations \citep{Buckle09}. All intensities presented in this paper are given in units of main beam temperature ($T_\text{MB}$), unless otherwise stated. The system temperatures for the $^{12}$CO, $^{13}$CO, and C$^{18}$O observations were roughly 600 K, 1000 K, and 1500 K, respectively. The effective integration time per spectrum in the final maps are 19.2 s for the $^{12}$CO observations, 258 s for the $^{13}$CO and C$^{18}$O observations towards IRDC C, and 224 s for the $^{13}$CO and C$^{18}$O observations towards IRDC F. The pointing accuracy of the JCMT is generally considered to be at worst of the order of a few arcseconds \citep{Buckle09}. All observations were conducted on May 8th, 2013 and further details of the observational setups are given in Table \ref{table:JCMT setup}. Relevant molecular data are given in Table \ref{table:molecular data}

\begin{table*}
\begin{minipage}{\textwidth}
\caption{JCMT Observation Setup}
\begin{center}
\begin{tabular}{cccccc}
\hline
\hline
IRDC & Lines Observed & RA & Dec & Off RA & Off Dec. \\
 & & (h:m:s) & ($^\circ$:\arcmin:\arcsec) & (h:m:s) & ($^\circ$:\arcmin:\arcsec) \\
(1) & (2) & (3) & (4) & (5) & (6) \\
\hline
C (G028.37+00.07) & $^{12}$CO (3-2), $^{13}$CO (3-2), C$^{18}$O (3-2) & 18:42:47.29 & -4:04:06.1 & 18:43:19.451 & -4:33:26.94 \\
F (G034.43+00.24) & $^{12}$CO (3-2), $^{13}$CO (3-2), C$^{18}$O (3-2) & 18:53:17.35 & 1:26:34.5  & 18:51:46.073 & 1:35:41.35 \\
\hline
\end{tabular}
\tablefoot{Column 1 gives the name of the IRDC observed while Col.~2 gives the transitions observed. Columns 3 and 4 are the right ascension and declination (J2000), respectively, of the map centers. The right ascension and declination of the off position are given in Cols.~5 and 6. }
\label{table:JCMT setup}
\end{center}
\end{minipage}
\end{table*}

\begin{table*}
\begin{minipage}{\textwidth}
\caption{Molecular Data}
\begin{center}
\begin{tabular}{ccccc}
\hline
\hline
Transition & $\nu$ & $E_{\text{u}}$ & $n_{\text{crit}}$ & $A$ \\
 & (GHz) & (K) & (10$^4$ cm$^{-3}$) & (10$^{-6}$ s$^{-1}$) \\
(1) & (2) & (3) & (4) & (5) \\
\hline
$^{13}$CO (2-1) & 220.39868 & 15.87 & 8.4 & 6.038 \\
C$^{18}$O (2-1) & 219.56035 & 15.81 & 8.4 & 6.011 \\
$^{12}$CO (3-2) & 345.79599 & 33.19 & 3.8 & 2.497 \\
$^{13}$CO (3-2) & 330.58797 & 31.73 & 3.3 & 2.181 \\
C$^{18}$O (3-2) & 329.33055 & 31.61 & 3.3 & 2.172 \\
\hline
\end{tabular}
\tablefoot{Column 1 gives the transition observed and Col.~2 gives the frequency of the observed transition. Column 3 gives the energy, in units of Kelvin, of the upper level of the transition and Col.~4 gives the critical density of the transition. Column 5 gives the Einstein $A$ coefficient for the transition. All molecular data are based upon data in the Leiden Atomic and Molecular Database (LAMDA; \citealt{Schoier05}). Data in the LAMDA partially come from the National Aeronautics and Space Administration (NASA) Jet Propulsion Laboratory (JPL) database \citep{Pickett98} and the Cologne Database for Molecular Spectroscopy \citep{Muller01, Muller05}.}
\label{table:molecular data}
\end{center}
\end{minipage}
\end{table*}

The data was first passed through the default JCMT ORAC-DR pipeline \citep{Cavanagh08}. The C$^{18}$O observations of IRDC C were initially rejected by the pipeline because the system temperature was too large. These C$^{18}$O, IRDC C data were thus manually processed by checking the data for noise spikes, combining the individual time series observations into a data cube and then finally fitting a linear baseline to the data. The C$^{18}$O spectra, while slightly noisy, appear by eye to still be viable and thus, are included in further analysis. 

All reduction of this JCMT data was done using the Starlink data reduction package. To improve the signal to noise ratio, all data sets were smoothed to a velocity resolution of 0.5 km s $^{-1}$. A combination of Gaussian profiles was then fit to each spectrum using the Figaro fitgauss Gaussian fitting command. Each spectrum was manually inspected and the number of components required to fit each spectrum was determined by eye (see below the criteria used to determine whether a velocity component was detected or not). 
 
In order to select only the emission likely to come from the IRDCs of interest, and not other structures along the line of sight, only components with central velocities with respect to the local standard of rest (LSR) between 49 and 65 km s$^{-1}$ for IRDC F and between 73 and 85 km s$^{-1}$ for IRDC C are considered as possible detections. These velocity extents are also chosen to match the velocity range over which emission is detected in the $^{12}$CO J = 8 $\rightarrow$ 7 and 9 $\rightarrow$ 8 transitions by the {\it Herschel Space Observatory} (Paper I), in order to facilitate the creation and analysis of spectral line energy distributions for these regions in Paper III of this series (Pon et al. in preparation). While there is some $^{12}$CO J = 3 $\rightarrow$ 2 emission outside of these velocity ranges, the majority of the $^{12}$CO emission and all of the clearly detected  $^{13}$CO and C$^{18}$O emission lie within these ranges. The emission outside of these ranges is likely due to CO emission from unrelated clouds at different velocities along the line of sight, as we do not believe we have detected any transitions other than the three CO isotopologue lines. 

For spectra with only one obvious component, the best fit is considered a detection only if the integrated intensity of the line is greater than three times the uncertainty of the integrated intensity determined by the Gaussian fitting command and greater than three times the uncertainty calculated via the equation
\begin{equation}
dI = RMS \times \sqrt{FWHM \times \delta v},
\label{eqn:rms}
\end{equation}
where $dI$ is the uncertainty in the integrated intensity, $RMS$ is the root mean square of the baseline, $FWHM$ is the full width at half maximum of the line, and $\delta v$ is the velocity resolution. If two or more components are apparent, each component is evaluated individually and detections are required to have an integrated intensity three times larger than the uncertainty calculated via Equation \ref{eqn:rms}. Since Equation \ref{eqn:rms} can produce small uncertainties for unrealistically narrow line fits, we also adopt the very loose constraint that all fits must have an uncertainty from the Gaussian fitting command less than ten times the derived integrated intensity. We set such a high tolerance for the uncertainty from the Gaussian fitting command because these uncertainties can become quite large when components are partially blended. Because of this increase in the Gaussian fitting command uncertainty, based upon the uncertainty of how to divide the observed flux between multiple components, we choose to use Equation \ref{eqn:rms} for the integrated intensity uncertainty, rather than the value from the Gaussian fitting command. For all spectra, the uncertainties in the FWHM and central velocity are taken to be the relevant uncertainties given by the Gaussian fitting command. 

An analysis of the IRDC F $^{13}$CO data set using the automated fitting routine of \citet{Henshaw14} produces similar results to the by-eye fitting, such that we have reasonable confidence in the validity of this by-eye method of selecting components. While the data have sufficient signal to noise to apply a moment analysis, it is clear that multiple components are common and many of these components are badly blended. As discussed in \citet{Henshaw14}, moment analysis can provide misleading results when there are multiple blended components, as an increase in the second moment can be caused by either an increase in the intrinsic FWHM of a line or by an increase in the separation between the central velocities of two partially blended lines. As such, we choose to rely on the Gaussian fits to the data, rather than a moment analysis. Multicomponent Gaussian fitting has been shown to produce good results in other IRDCs (e.g., \citealt{JimenezSerra14}). 

\begin{figure}
   \centering
   \includegraphics[width=3.5in]{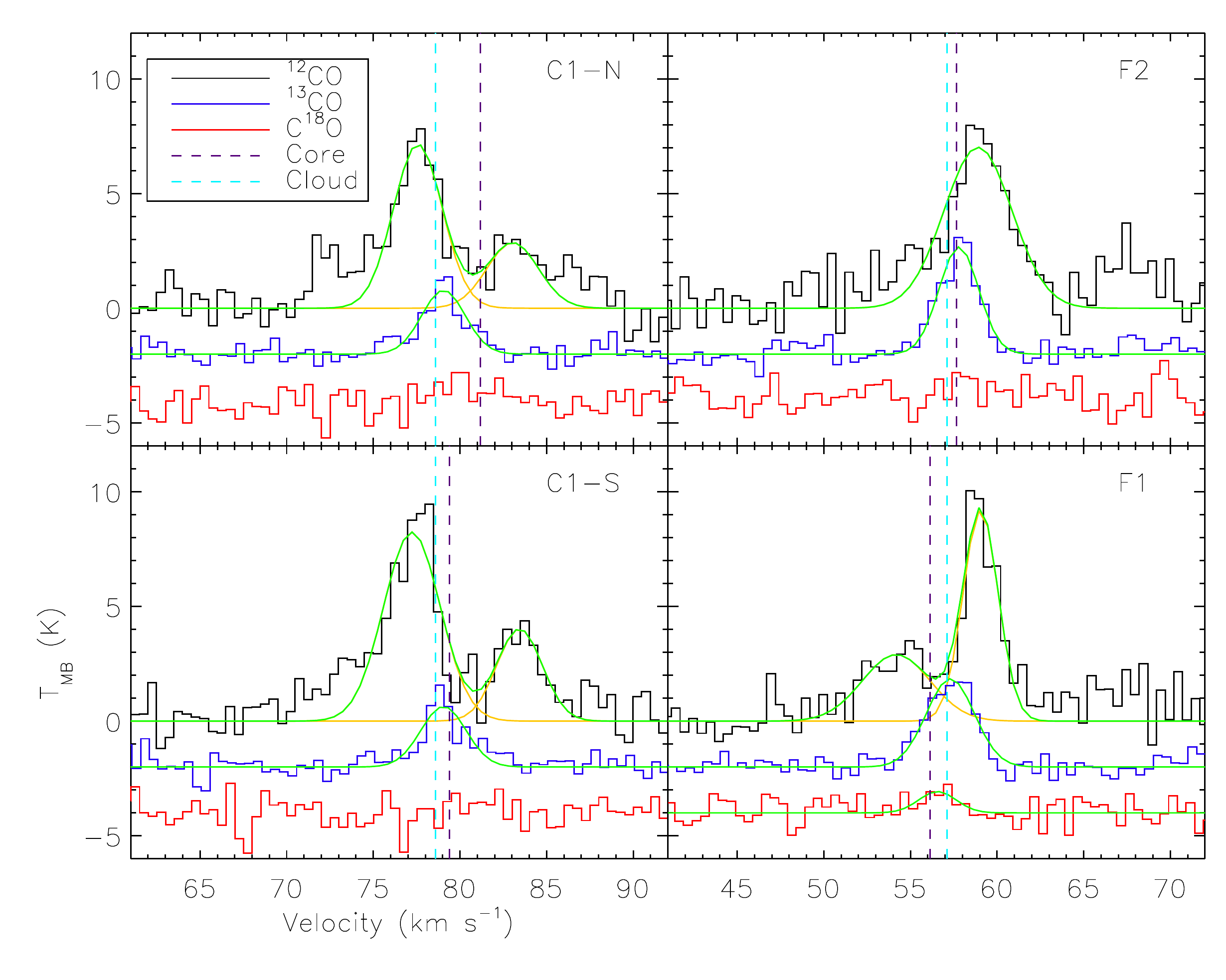}
   \caption{JCMT spectra towards the central positions of the C1-N, C1-S, F1, and F2 cores, with the cores labeled in the top right hand corner of the spectral boxes. All spectra are for the J = 3 $\rightarrow$ 2 transition, with the black showing the $^{12}$CO data, the blue showing the $^{13}$CO data shifted down by 2 K, and the red showing the C$^{18}$O data shifted down by 4 K. For each spectrum, the green lines show the cumulative best fit of all of the components within the velocity range defined for the appropriate cloud, while the yellow line shows the individual components contributing to the fit. The vertical, dashed, dark purple lines give the central velocity of the N$_2$D$^+$ J = 3 $\rightarrow$ 2 line detected towards the cores \citep{Tan13}, while the dashed, light blue lines give the central velocities of the parent IRDCs \citep{Simon06}. }
   \label{fig:sample}
\end{figure}

The spectra towards the central locations of the C1-N, C1-S, F1, and F2 cores are shown in Fig.~\ref{fig:sample}. This figure also shows the best fitting Gaussian components and the cumulative best fit for each spectrum. The full set of observed spectra and their corresponding fits are shown in Appendix \ref{appendix:jcmtfits}. Table \ref{table:JCMT fits} presents the extreme and average quantities from the Gaussian fitting of the JCMT data. 

\begin{table*}
\begin{minipage}{\textwidth}
\caption{JCMT Fits}
\begin{center}
\begin{tabular}{cccccccccc}
\hline
\hline
IRDC & Transition & $RMS_\text{mean}$ & $T_\text{MB,max}$ & $T_\text{MB,ave}$ & $I_\text{max}$ & $I_\text{ave}$ & $FWHM_\text{max}$ & $FWHM_\text{min}$ & $FWHM_\text{ave}$ \\
 & & (K) & (K) & (K) & (K km s$^{-1}$) & (K km s$^{-1}$) & (km s$^{-1}$) & (km s$^{-1}$) & (km s$^{-1}$) \\
(1) & (2) & (3) & (4) & (5) & (6) & (7) & (8) & (9) & (10) \\
\hline
C & $^{12}$CO(3-2) & 0.8 & 12.6 & 7.5 & 68.0 & 39.4 & 14.9 & 0.6 & 3.5 \\
C & $^{13}$CO(3-2) & 0.4 & 4.5 & 2.6 & 16.7 & 9.8 & 7.7 & 0.8 & 3.6 \\
C & C$^{18}$O(3-2) & 0.5 & 2.7 & 1.5 & 6.2 & 3.6 & 8.1 & 0.5 & 1.9 \\
F & $^{12}$CO(3-2) & 0.7 & 12.0 & 8.3 & 60.9 & 36.5 & 11.0 & 0.6 & 2.9 \\
F & $^{13}$CO(3-2) & 0.4 & 8.9 & 3.7 & 38.8 & 11.8 & 5.2 & 0.6 & 2.6 \\
F & C$^{18}$O(3-2) & 0.6 & 2.9 & 1.5 & 8.5 & 3.2 & 4.7 & 0.5 & 2.0 \\
\hline
\end{tabular}
\tablefoot{Column 1 gives the name of the IRDC observed with the JCMT while Col.~2 gives the transition observed. The average RMS value of the baseline is given in Col.~3. The maximum intensity in a map and the mean peak intensity of all spectra with detections in a map are given in Cols.~4 and 5, respectively. The maximum and mean of the sum of the integrated intensities of all of the components in one spectrum are given in Cols.~6 and 7, respectively. Columns 8, 9, and 10 give the maximum, minimum, and mean FWHM of all components, respectively. For all values given, the uncertainty is in the last digit.}
\label{table:JCMT fits}
\end{center}
\end{minipage}
\end{table*}

\begin{figure*}
   \centering
   \begin{subfigure}[b]{0.33\textwidth}
      \centering
      \includegraphics[width=2.6in]{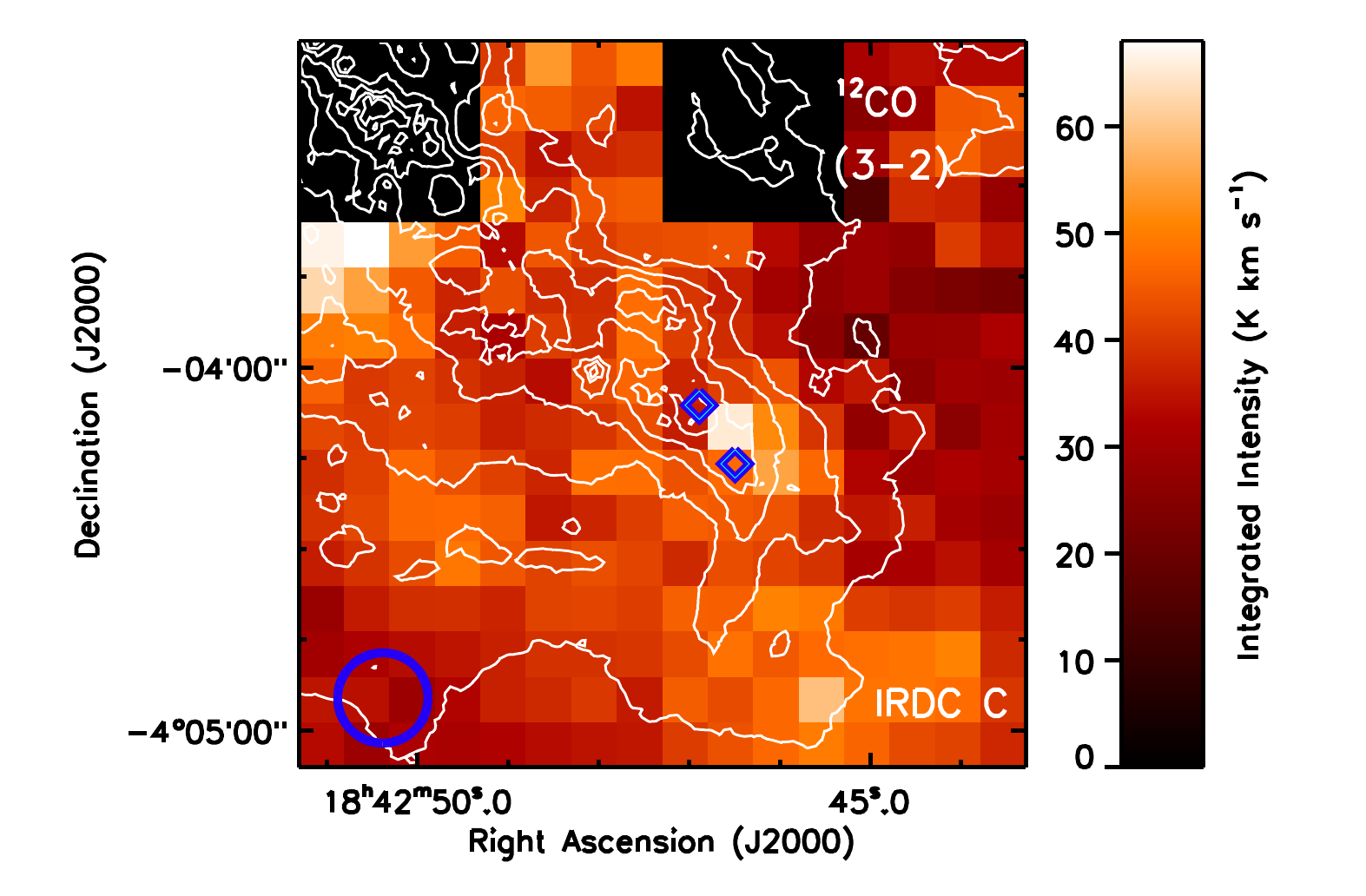}
   \end{subfigure}%
  \begin{subfigure}[b]{0.33\textwidth}
     \centering
     \includegraphics[width=2.6in]{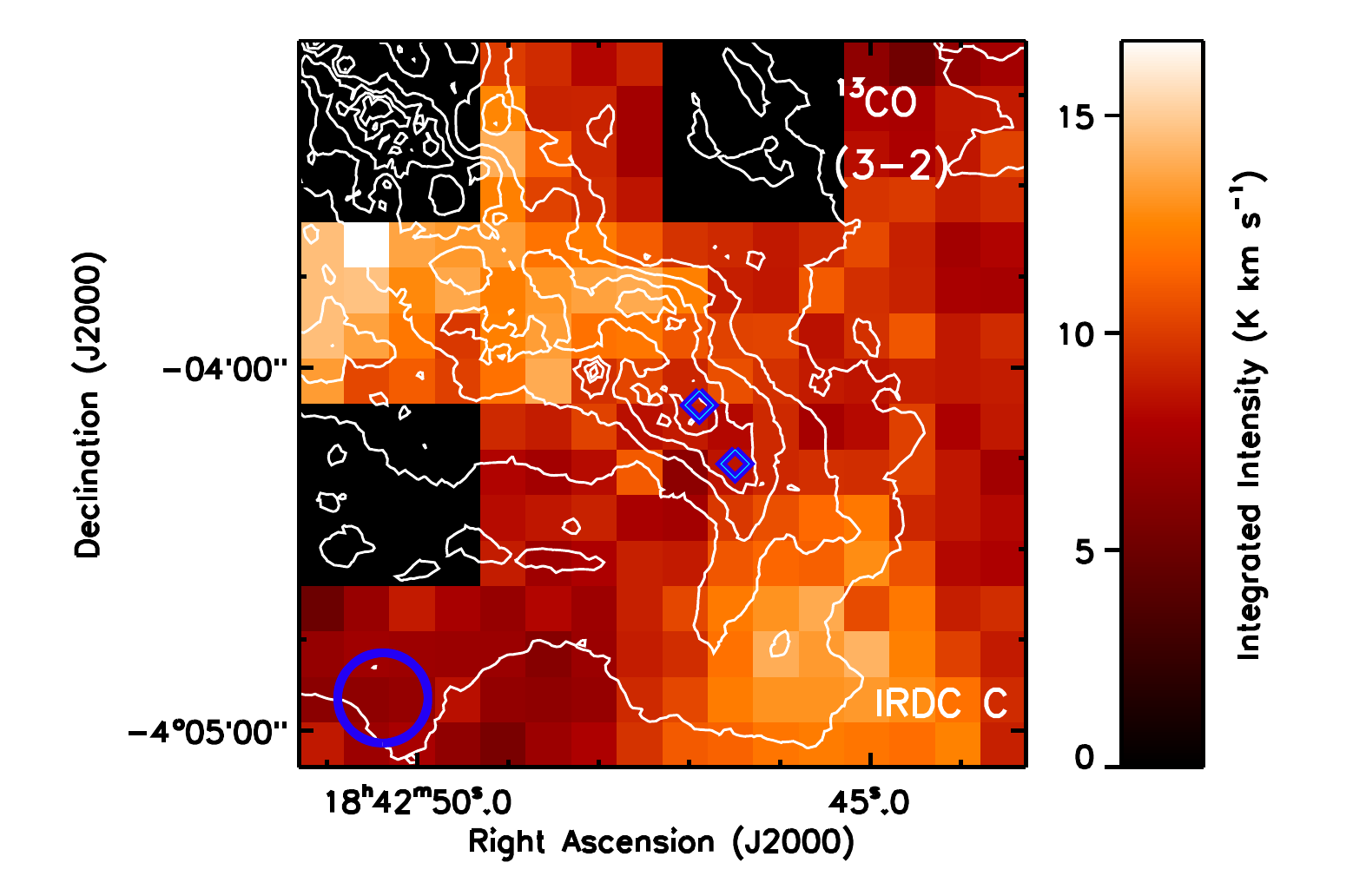}
   \end{subfigure}
   \begin{subfigure}[b]{0.33\textwidth}
      \centering
     \includegraphics[width=2.6in]{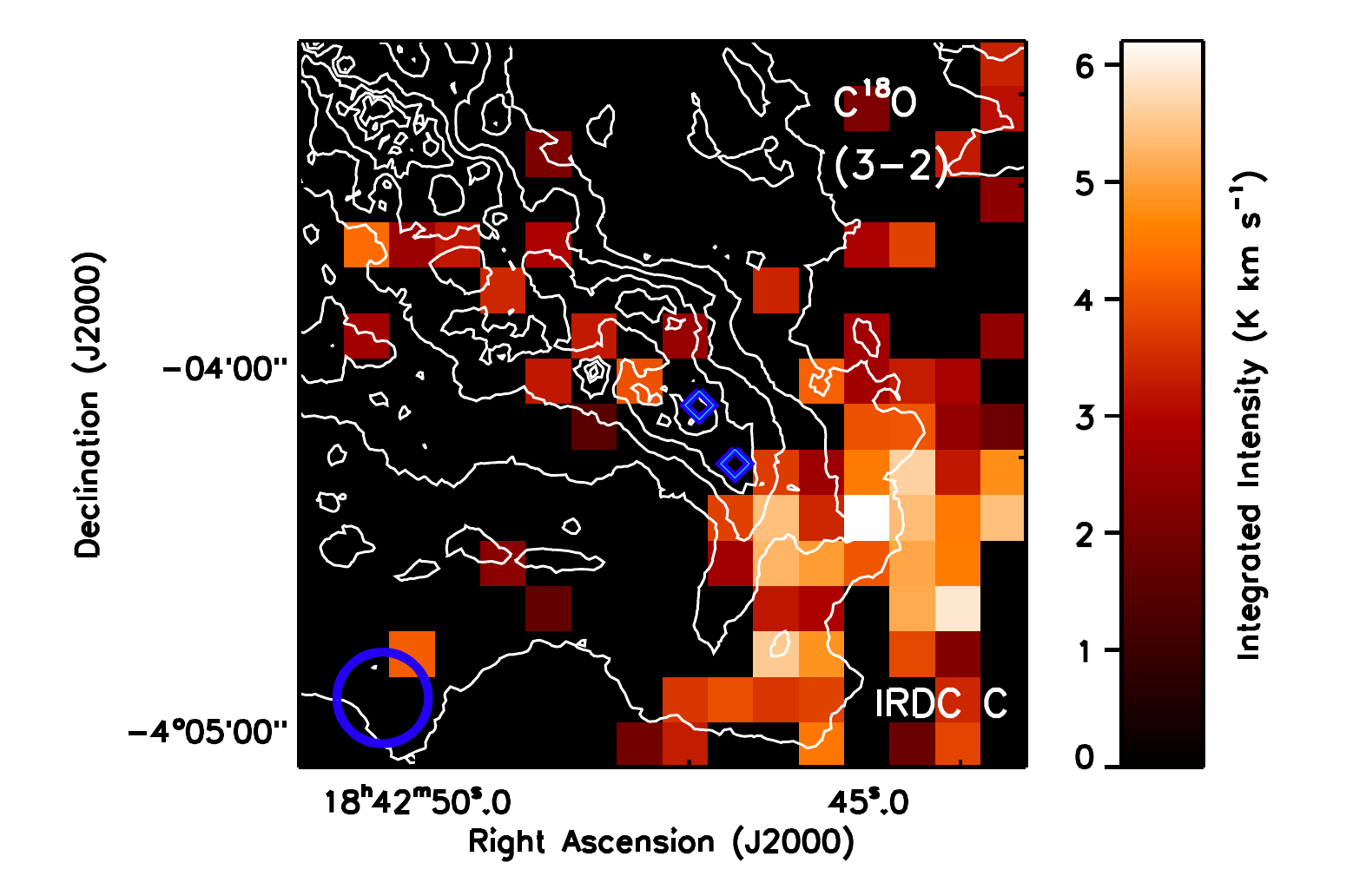}
   \end{subfigure}
   \caption{Total integrated intensities for all detected components of the $^{12}$CO (left), $^{13}$CO (center), and C$^{18}$O (right) J = 3 $\rightarrow$ 2 lines towards IRDC C by the JCMT (in color scale). The contours give the mass surface density derived by \citet{Butler12} and are the same as in Fig.~\ref{fig:sigmaradec}. The blue diamonds give the central locations of the C1-N and C1-S cores, with the top left core in the subfigures being C1-N. The blue circle shows the HPBW of the observations.}
   \label{fig:ctotalint}
\end{figure*}

\begin{figure*}
   \centering
   \begin{subfigure}[b]{0.33\textwidth}
      \centering
	   \includegraphics[width=2.6in]{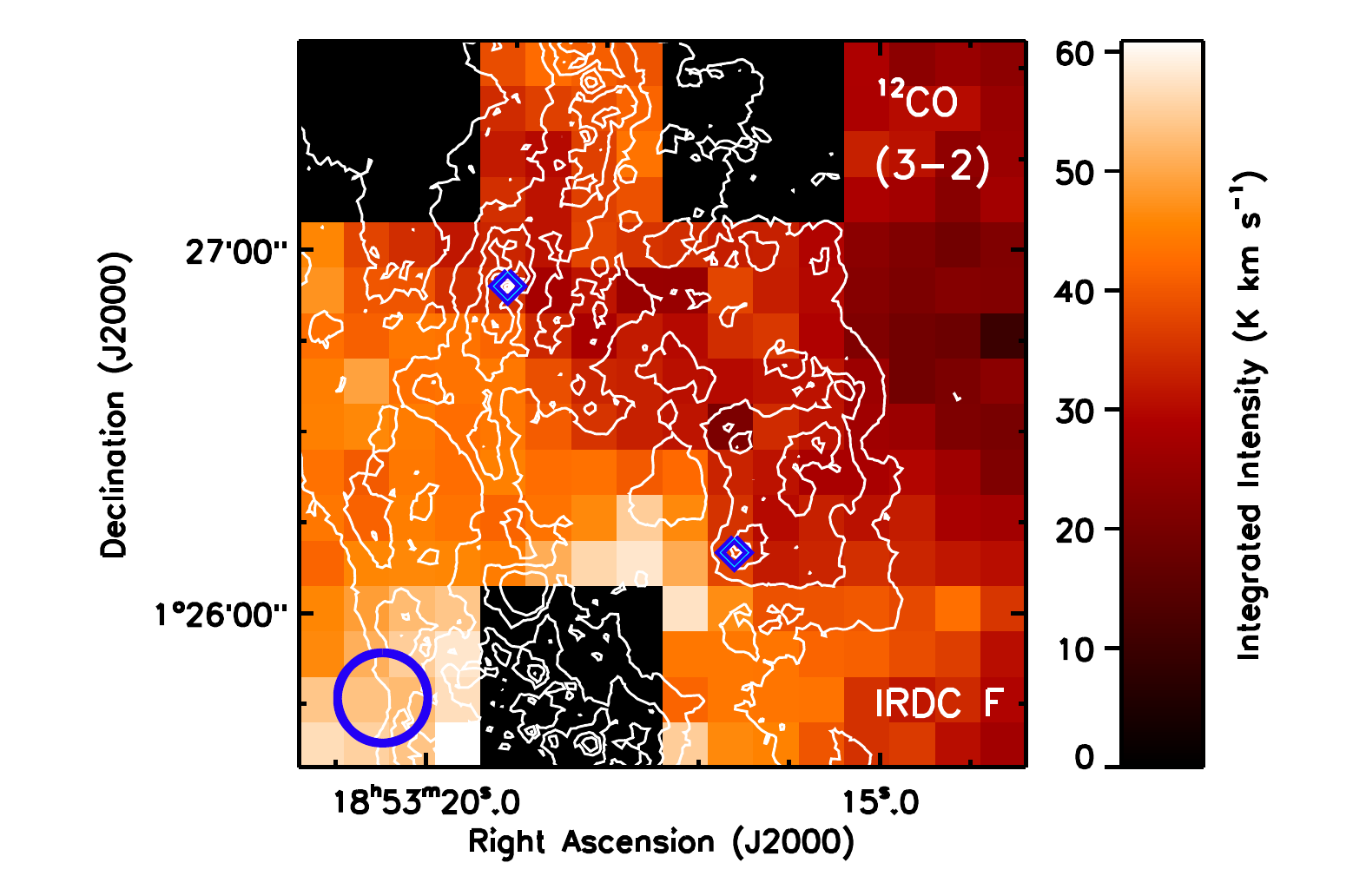}
   \end{subfigure}%
  \begin{subfigure}[b]{0.33\textwidth}
     \centering
	 \includegraphics[width=2.6in]{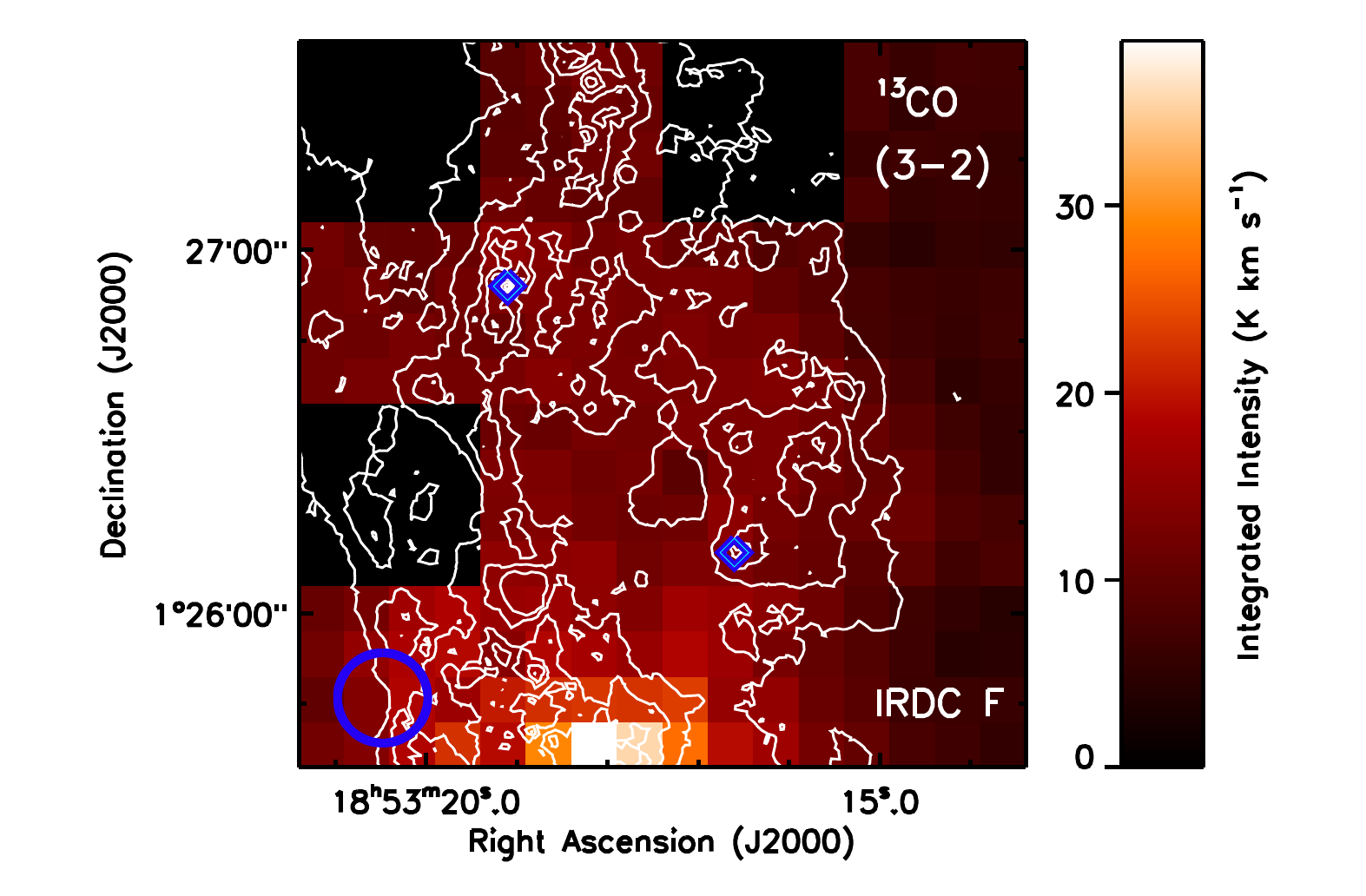}
   \end{subfigure}
   \begin{subfigure}[b]{0.33\textwidth}
      \centering
	 \includegraphics[width=2.6in]{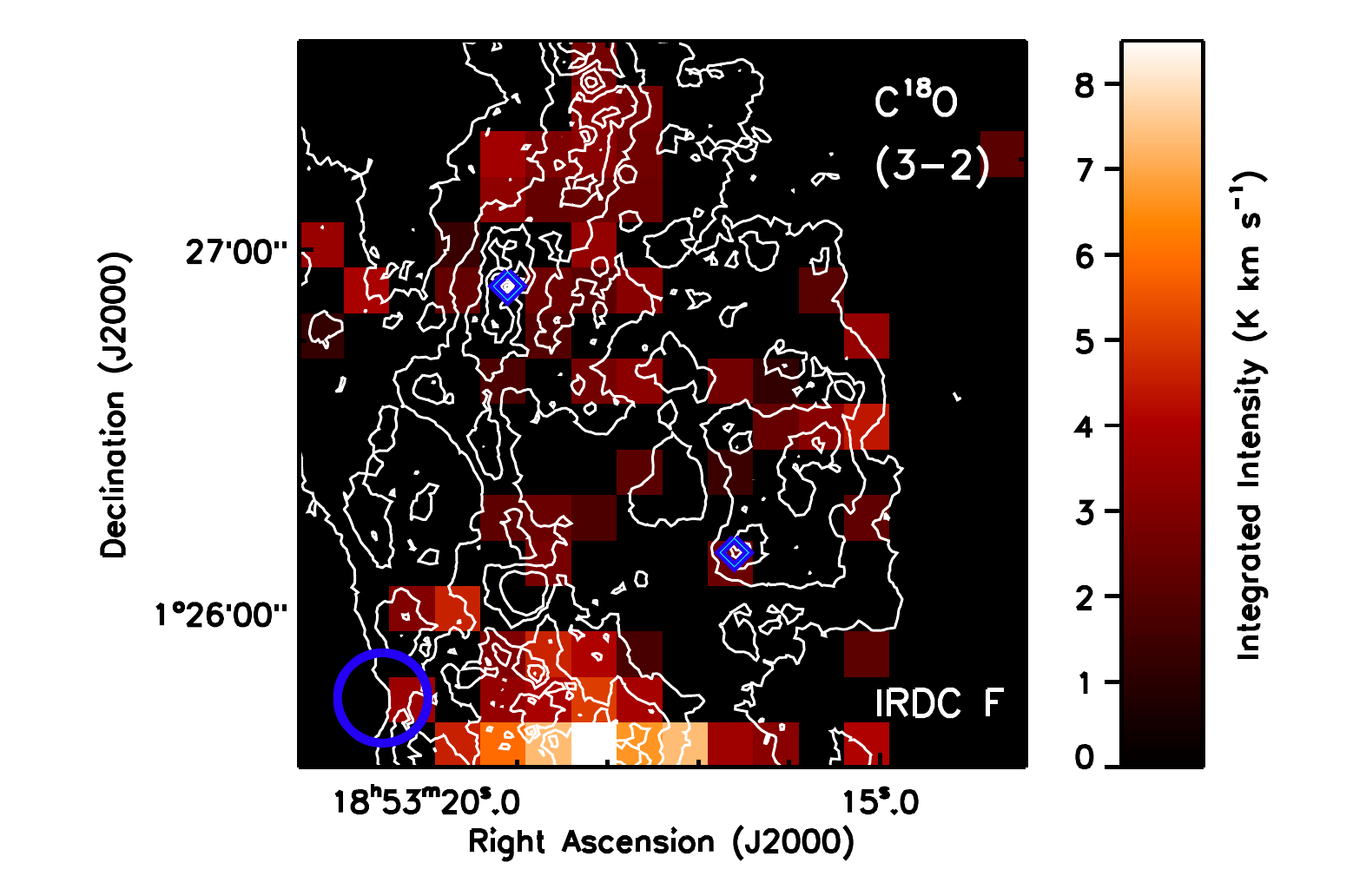}
   \end{subfigure}
   \caption{Total integrated intensities for all detected components of the $^{12}$CO (left), $^{13}$CO (center), and C$^{18}$O (right) J = 3 $\rightarrow$ 2 lines towards IRDC F by the JCMT (in color scale). The contours give the mass surface density derived by \citet{Butler12} and are the same as in Fig.~\ref{fig:sigmaradec}. The blue diamonds give the central locations of the F1 and F2 cores (right and left in the subfigures, respectively). The blue circle shows the HPBW of the observations.}
   \label{fig:ftotalint}
\end{figure*}

Figures \ref{fig:ctotalint} and \ref{fig:ftotalint} show the integrated intensity sum of all detected components for the three lines observed with the JCMT towards IRDCs C and F. The overlaid contours are the mass surface densities derived by \citet{Butler12}. 

The CO High-Resolution Survey (COHRS) is a JCMT survey of the Galactic plane in the $^{12}$CO J = 3 $\rightarrow$2 line, which has covered IRDCs F and C, as well as our off position for IRDC C \citep{Dempsey13}. We find that there is relatively little $^{12}$CO J = 3 $\rightarrow$ 2 emission at the velocities associated with C1 in our off position, such that it is unlikely that our lines are missing significant flux due to emission in the off position. We find reasonable agreement in the integrated intensities between our data set and that of the COHRS, with the sum of channels between 51 km s$^{-1}$ and 64 km s$^{-1}$ producing integrated intensities that agree to within 20\% for the positions closest to the F1, F2, and C1 clump centers. Similarly, \citet{Sanhueza10} observed the $^{12}$CO 3 $\rightarrow$ 2 and $^{13}$CO 3 $\rightarrow$ 2 transitions across IRDC F using the Atacama Pathfinder Experiment (APEX) telescope. They produced maps with 20 arcsecond pixel spacing for the 18 arcsecond beam of APEX. For the locations closest to the F1 and F2 core centers, the integrated intensities obtained by summing the channels between 51 km s$^{-1}$ and 64 km s$^{-1}$ produce integrated intensities within 15\% of those obtained from our JCMT observations, even with the slightly different beam sizes of the JCMT and APEX data.  

\subsection{IRAM 30 Meter Data}
\label{Fontani}

As part of the \citet{Fontani15Busquet} and \citet{Fontani15Caselli} survey, single pointing observations were obtained with the Institut de Radioastronomie Millim\'{e}trique (IRAM) 30 meter telescope, in early 2013, towards the central positions of the C1, F1, and F2 clumps given by \citet{Butler12}, corresponding roughly to the locations of the F1, F2, and C1-N cores. Both the $^{13}$CO and C$^{18}$O J = 2 $\rightarrow$ 1 transitions were observed. The IRAM 30 meter telescope has a beam size of approximately 11 arcsec at these frequencies and a main beam efficiency of 0.61. The resulting spectra have velocity resolutions of 0.27 km s$^{-1}$. The data have been fully reduced by Fontani et al. and the technical details of the observations can be found in \citet{Fontani15Busquet}. 

\begin{figure*}
   \centering
   \includegraphics[width=6.5in]{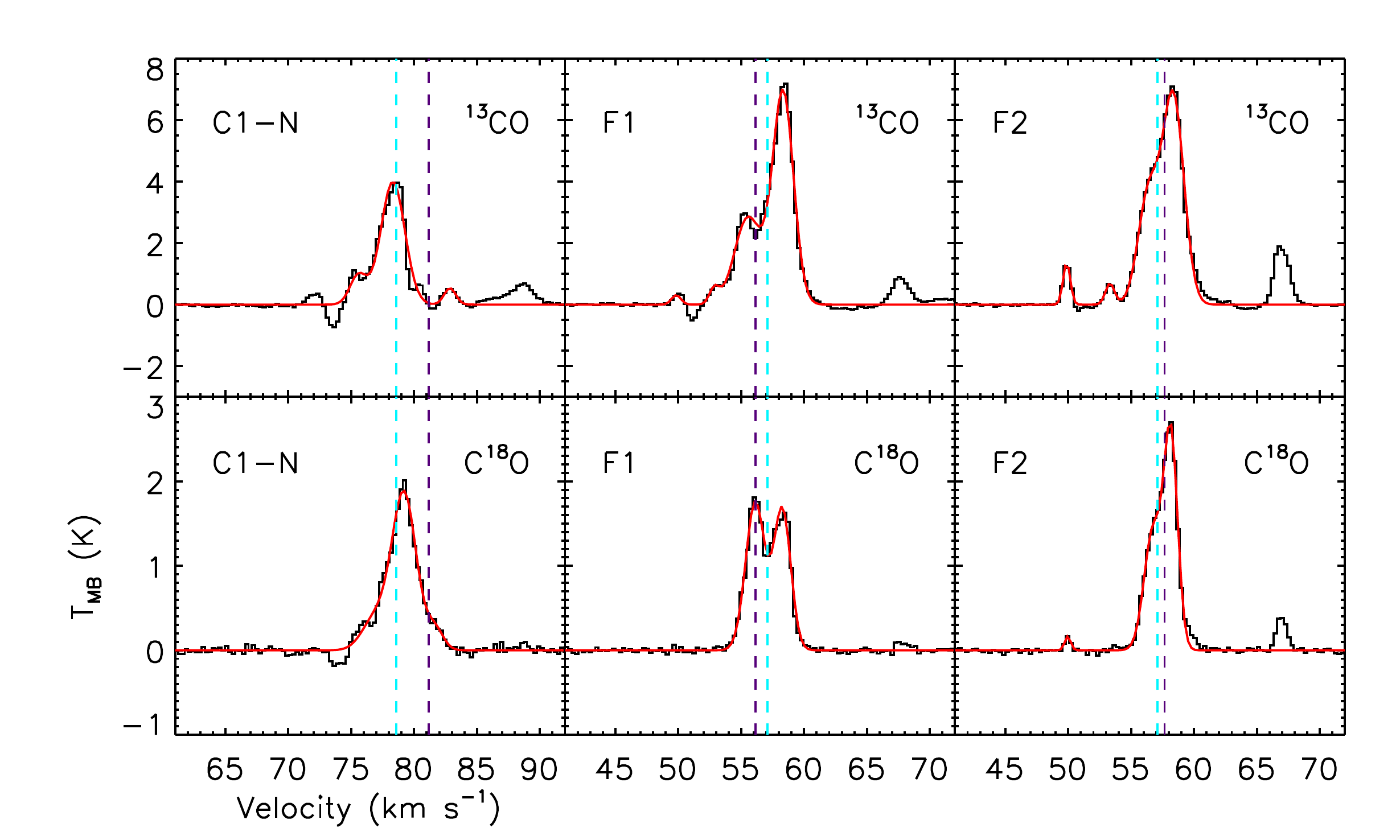}
   \caption{IRAM 30m $^{13}$CO and C$^{18}$O J = 2 $\rightarrow$ 1 spectra towards the C1-N, F1, and F2 cores. The red lines show the best fits to the data. The cores are labeled in the top left of each spectral window and the isotopologue in the top right. The vertical, dashed, dark purple lines give the central velocity of the N$_2$D$^+$ J = 3 $\rightarrow$ 2 line detected towards the cores \citep{Tan13} while the dashed, light blue lines gives the central velocities of the parent IRDCs \citep{Simon06}. }
   \label{fig:iramfits}
\end{figure*}

All six spectra show multiple components, consistent with the complex velocity structure seen in the JCMT observations. Using the built in Gaussian fitting routine in the GILDAS Continuum and Line Analysis Single-dish Software (CLASS) package, Gaussians were fit to the J = 2 $\rightarrow$ 1 data. The number of Gaussians required for each spectrum was determined by eye. The central velocities of detected components are required to lie within the same ranges as used for the JCMT data, 49 to 65 km s$^{-1}$ for IRDC F and 73 to 85 km s$^{-1}$ for IRDC C. The uncertainty in the integrated intensity for each component is determined via Equation \ref{eqn:rms} and the uncertainty in the total integrated intensity for a spectrum is taken to be the sum of the uncertainties of the components in that spectrum. All fitted components have integrated intensities well above five times their uncertainties. Figure \ref{fig:iramfits} shows the $^{13}$CO and C$^{18}$O J = 2 $\rightarrow$ 1 spectra towards the C1, F1, and F2 clumps. The red lines show the best fits to the data. Table \ref{table:IRAM} summarizes the key properties of the IRAM 30m observations.

\begin{table*}
\begin{minipage}{\textwidth}
\caption{IRAM 30m data}
\begin{center}
\begin{tabular}{ccccccc}
\hline
\hline
Core & Transition & RA (J2000) & Dec (J2000) & $I$ & $dI$ & $RMS$\\
 & & (h:m:s) & ($^\circ$:\arcmin:\arcsec) & (K km s$^{-1}$) & (K km s$^{-1}$) & (K)\\
(1) & (2) & (3) & (4) & (5) & (6) & (7) \\
\hline
C1-N & $^{13}$CO (2-1) & 18:42:46.95 & -4:04:08.5 & 11.80 & 0.06 & 0.02 \\
F1 & $^{13}$CO (2-1) & 18:53:16.46 & 1:26:09.9 & 22.85 & 0.08 & 0.02 \\
F2 & $^{13}$CO (2-1) & 18:53:19.06 & 1:26:52.6 & 23.39 & 0.07 & 0.02 \\
C1-N & C$^{18}$O (2-1) & 18:42:46.95 & -4:04:08.5 & 5.92 & 0.08 & 0.03 \\
F1 & C$^{18}$O (2-1) & 18:53:16.46 & 1:26:09.9 & 6.16 & 0.05 & 0.02 \\
F2 & C$^{18}$O (2-1) & 18:53:19.06 & 1:26:52.6 & 6.03 & 0.05 & 0.02 \\
\hline
\end{tabular}
\tablefoot{Column 1 gives the name of the core observed while Col.~2 gives the transition observed. Columns 3 and 4 are the right ascension and declination, respectively, of the observation. The sum of the integrated intensities of all detected components is given in Col.~5. The uncertainty in this total integrated intensity is calculated as the sum of the uncertainties of the individual components and is given in Col.~6. The RMS of the baseline is given in Col.~7.}
\label{table:IRAM}
\end{center}
\end{minipage}
\end{table*}

Both spectra towards the C1 clump show a small absorption feature just blue-wards of the main line, which may be indicative of outflowing gas, but is more likely due to emission at that velocity in the off position used, as the observations were conducted in wobbler switching mode and an emission line centered at roughly 100 km s$^{-1}$ also shows a strong blue-shifted absorption feature.

\section{Temperature}
\label{temperature}

	To derive CO column densities and depletion factors from the $^{13}$CO observations in Sect.~\ref{depletion}, the excitation temperature of the $^{13}$CO lines is required. Prior gas kinetic temperature estimates for IRDC F range from 5 K to 20 K (\citealt{RomanDuval10, Tan13, Foster14}; Barnes et al.~in prep.), while temperatures from 8 to 30 K have been suggested for IRDC C \citep{Wang08, RomanDuval10, Wang12, Tan13}. In this section, we attempt to constrain the excitation temperature of the $^{13}$CO lines from the JCMT and IRAM observations described in Sect.~\ref{observations}.

\begin{figure*}
   \centering
   \begin{subfigure}[b]{0.49\textwidth}
     \centering
       \includegraphics[width=4in]{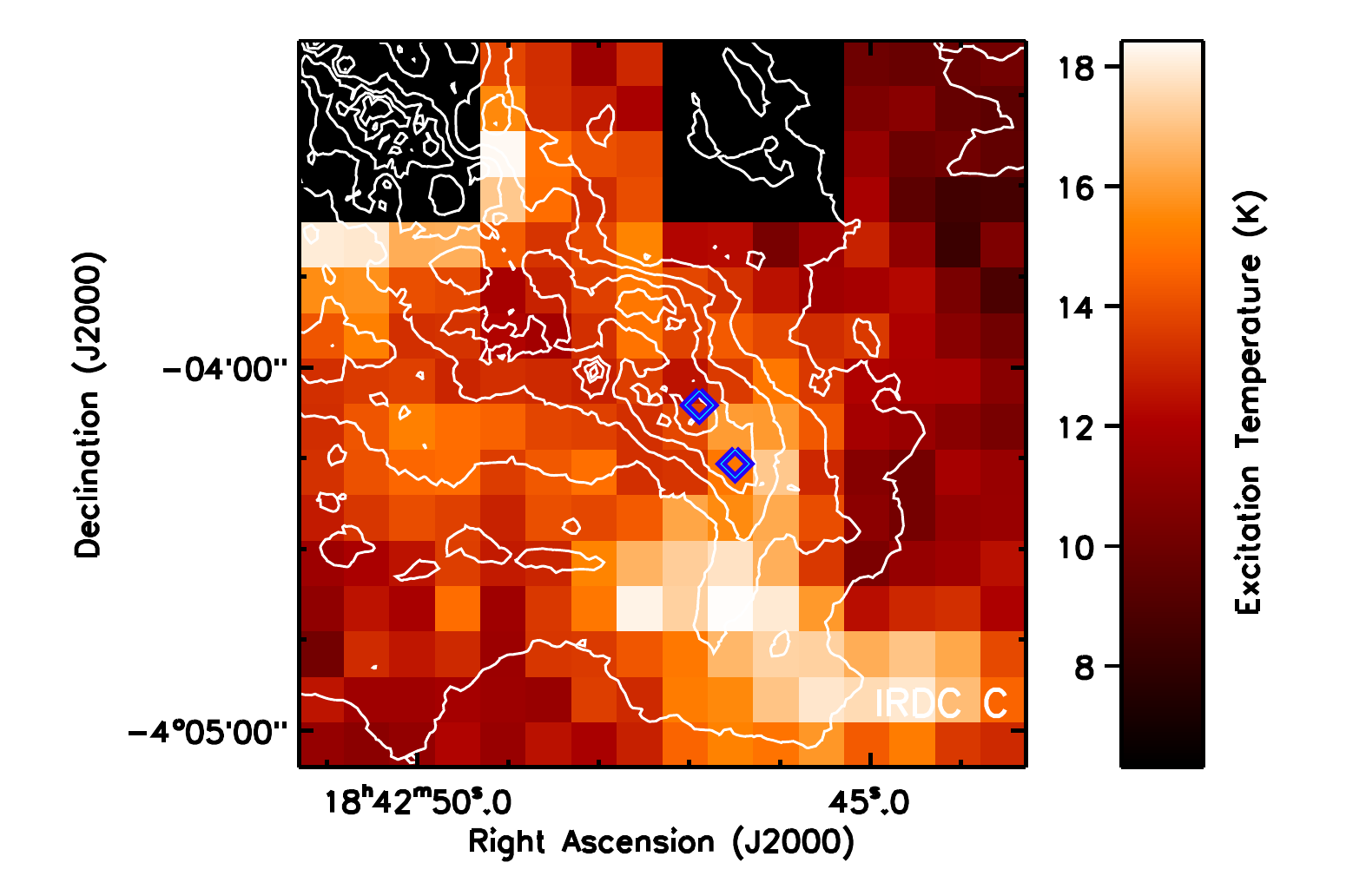}
   \end{subfigure}
   \begin{subfigure}[b]{0.49\textwidth}
      \centering
      \includegraphics[width=4in]{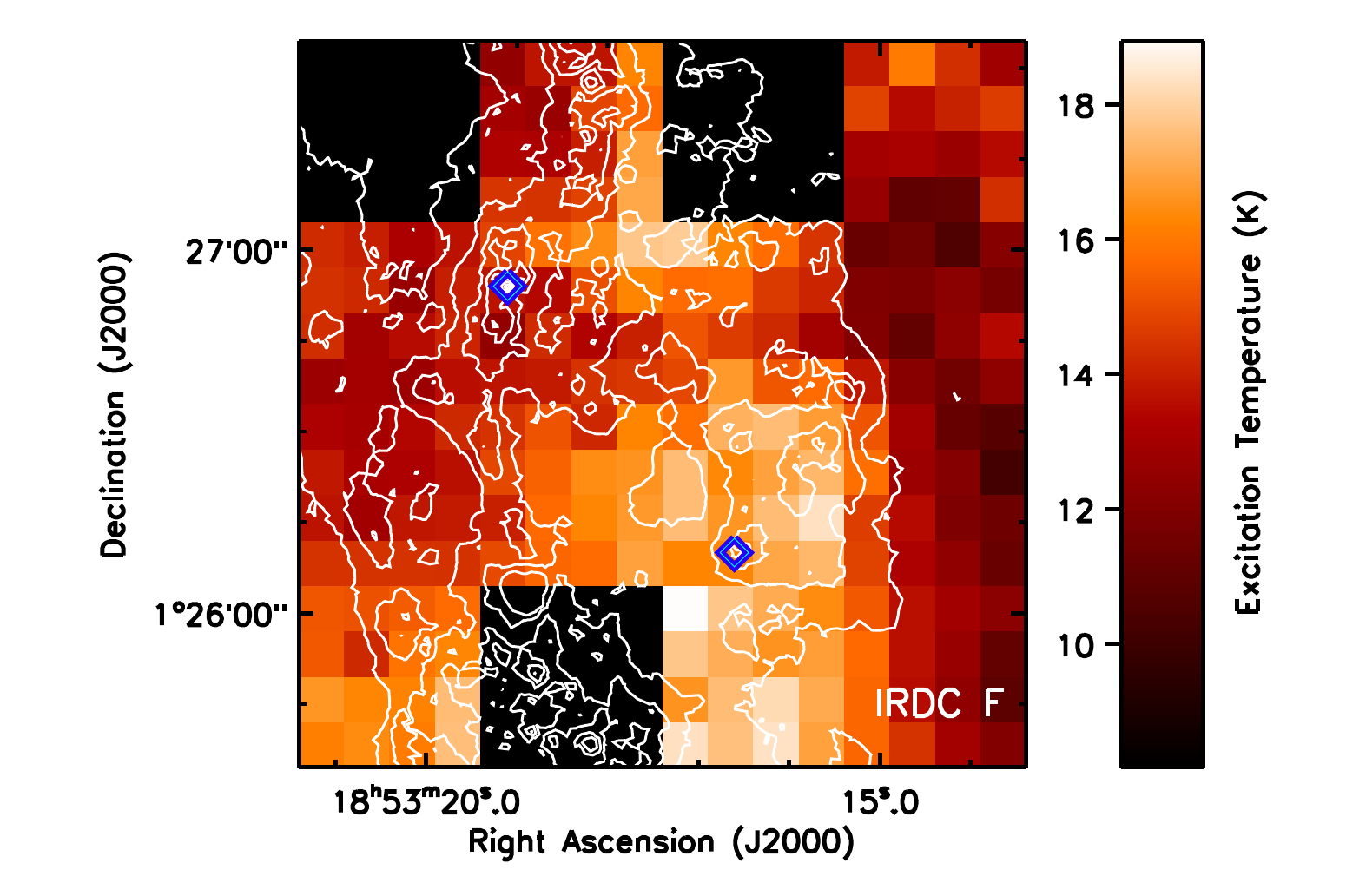}
   \end{subfigure}
   \caption{Excitation temperatures based on the $^{12}$CO J = 3 $\rightarrow$ 2 line towards IRDC C (left) and IRDC F (right). The contours give the mass surface density derived by \citet{Butler12} and are the same as in Fig.~\ref{fig:sigmaradec}. The blue diamonds give the central locations of the C1-N, C1-S, F1, and F2 cores.}
   \label{fig:excitation}
\end{figure*}

The $^{12}$CO J = 3 $\rightarrow$ 2 line should be highly optically thick and thus, the excitation temperature of the J = 3 and J = 2 levels should be approximately equal to the kinetic temperature of the $^{12}$CO gas. For an optically thick line, where $\text{e}^{-\tau} \rightarrow 0$ ($\tau$ being the optical depth), the excitation temperature is related to the peak observed main beam temperature by 
\begin{equation}
T_\text{ex} = T_\text{0} \left[\ln\left(\frac{T_\text{0}}{T_\text{MB} + \frac{T_\text{0}}{\mbox{exp}\left(T_\text{0}/T_\text{bg}\right)-1}} + 1\right)\right]^{-1},
\label{eqn:tex}
\end{equation}
where $T_\text{ex}$ is the excitation temperature of the emitting gas, $T_\text{MB}$ is the peak main beam temperature of the line, $T_\text{bg}$ is the 2.7 K background temperature, and $T_\text{0}$ is the temperature corresponding to the energy difference between the upper and lower states of the transition. The excitation temperature of the $^{12}$CO J = 3 $\rightarrow$ 2 line is of the order of 10 to 20 K throughout the JCMT observed fields, with a mean temperature of 13.4 K in IRDC C and 14.6 in IRDC F. Figure \ref{fig:excitation} shows the derived excitation temperatures for the two mapped regions. Any self-absorption in the $^{12}$CO spectra should lower the observed peak intensity, which would cause underpredictions of the temperature within the clouds. The uncertainties in the excitation temperatures derived from the $^{12}$CO J = 3 $\rightarrow$ 2 line are found by evaluating Equation \ref{eqn:tex} with the peak intensity increased by the RMS of the baseline.

In IRDC F, higher temperatures are detected in the southern half of the map, probably due to the presence of an active star-forming clump at the southern edge of the map, which is responsible for the increased CO emission seen in Fig.~\ref{fig:ftotalint}. The MM7 water maser in the northern portion of the IRDC F map also seems to be associated with slightly elevated gas temperatures. In IRDC C, there is an increase in temperature in the northeast corner of the map, towards the water maser in this corner of the map. While these three temperature increases seem to be clearly related to embedded sources, there is an additional temperature increase to the southwest of C1-N and C1-S that is not clearly associated with any further embedded sources. This temperature increase in the southwest corner of the IRDC C map may be evidence for external illumination of the cloud or perhaps extra heating from accretion onto the main IRDC C cloud. 

Starless cores are expected to have cooler interiors than exteriors, as the incident UV field is absorbed by the outer layers of gas (e.g., \citealt{Evans01, Zucconi01}). Since $^{13}$CO and C$^{18}$O are less abundant isotopologues than $^{12}$CO, the $\tau = 1$ surface for the $^{13}$CO and C$^{18}$O J = 3 $\rightarrow$ 2 transitions should occur further into the cloud than the  $^{12}$CO J = 3 $\rightarrow$ 2 $\tau = 1$ surface. Equivalently, the $^{13}$CO and C$^{18}$O J = 3 $\rightarrow$ 2 transitions should be less optically thick than the $^{12}$CO J = 3 $\rightarrow$ 2 transition. Therefore, the $^{13}$CO and C$^{18}$O lines may trace slightly cooler gas, closer to the well-shielded interiors of the IRDCs, than the $^{12}$CO line. 

Furthermore, while the excitation temperature of the $^{12}$CO J = 3 $\rightarrow$ 2 line is likely equal to the kinetic temperature of the gas, due to the very large ($\tau > 100$) optical depth of the line, the $^{13}$CO and C$^{18}$O J = 3 $\rightarrow$ 2 lines may be slightly subthermally excited (excitation temperatures less than the kinetic temperature). This is because the critical densities for the J = 3 $\rightarrow$ 2 transitions, of the order of 3 $\times 10^{4}$ cm$^{-3}$, are not significantly less than the expected densities of IRDCs \citep{Rathborne06, Tan14}, and the lines of the less abundant isotopologues are much less optically thick.

Towards the central locations of the F1, F2, and C1-N cores, the IRAM 30m observations of the $^{13}$CO and C$^{18}$O J = 2 $\rightarrow$ 1 lines allow for an independent measurement of the gas kinetic temperature and the relevant $^{13}$CO excitation temperatures, by comparing the ratios of the integrated intensities of these two lines to the $^{13}$CO J = 3 $\rightarrow$ 2 line. Due to the low signal to noise of the C$^{18}$O J = 3 $\rightarrow$ 2 line, we do not consider it in this calculation. 

We use the RADEX code \citep{Vandertak07} to calculate the emission coming from a uniform density, uniform temperature slab of material. The background temperature is set to 2.73 K and the H$_2$ density is set to $5 \times 10^4$ cm$^{-3}$ for C1-N and $2 \times 10^{4}$ cm$^{-3}$ for F1 and F2 \citep{Butler09}. The peak densities are likely higher in these cores \citep{Butler12, Tan13,Butler14}, but the lower density envelopes should be contributing significantly to the observed CO emission and be the main source of CO emission at the angular resolution of the IRAM and JCMT observations. That is, we believe these CO observations are likely tracing the bulk envelope around these cores, rather than just the dense interiors of these cores. For the $^{13}$CO J = 3 $\rightarrow$ 2 line, only one component is detected towards C1-N, F1, and F2, such that we set the FWHMs of the models to be equal to the observed FWHMs. For the J = 2 $\rightarrow$ 1 lines, there are two significant components in each core, except for the C1-N C$^{18}$O spectrum, where there are three significant components. We set the integrated intensity to be the sum of all of these components. For the lines with two dominant components, we set the FWHM in the RADEX model to be equal to half of the sum of the FWHM of the two components plus the difference in line centroids of the two components, thereby providing a rough estimate of the overall FWHM of the blended line. For the C1-N C$^{18}$O J = 2 $\rightarrow$ 1 line, we set the FWHM in RADEX to be equal to the centroid difference between the two outermost components, which are both much weaker than the central component. Given this approximate treatment of the multiple components seen in the J = 2 $\rightarrow$ 1 data, for this calculation, we adopt a conservative 25\% uncertainty in the integrated intensities of the lines. This 25\% level is based upon the central component in the C1-N spectra only contributing about 75\% of the total integrated intensity of the observed J = 2 $\rightarrow$ 1 lines. For this calculation, we also set the uncertainty for the J = 3 $\rightarrow$ 2 line to be 25\%, instead of being of the order of 2 to 5\%, so that the lines are equally weighted. We assume that the lines fully fill the beam and do not adjust for the small beam size difference between the IRAM and JCMT data. 

To link the $^{13}$CO models to the C$^{18}$O RADEX models, we adopt the isotope ratios from \citet{Wilson94} of
\begin{eqnarray}
^{12}\mbox{C} / ^{13}\mbox{C} &=& 7.5 D_\text{GC}(\mbox{kpc}) + 7.6,\\
^{16}\mbox{O} / ^{18}\mbox{O} &=& 58.8 D_\text{GC}(\mbox{kpc}) + 37.1,
\end{eqnarray}
where $D_\text{GC}$ is the Galactocentric radius (4.7 kpc and 5.8 kpc for IRDCs C and F, respectively). It is assumed that the ratios of the CO isotopologues are the same as the above isotope ratios, such that the $^{12}$CO / $^{13}$CO and $^{12}$CO / C$^{18}$O ratios are 43 and 313 in IRDC C, and 51 and 378 in IRDC F. 

We find the best fitting model by minimizing the sum of the squares of the intensity differences between the models and observations, scaled by the uncertainties of the integrated intensities. That is, minimizing
\begin{equation}
\chi^2 = \Sigma \left(\left[\frac{I_\text{model}}{I_\text{obs}} - 1\right] / 0.25\right)^2 .
\end{equation}

\begin{figure*}
   \centering
    \includegraphics[width=7in]{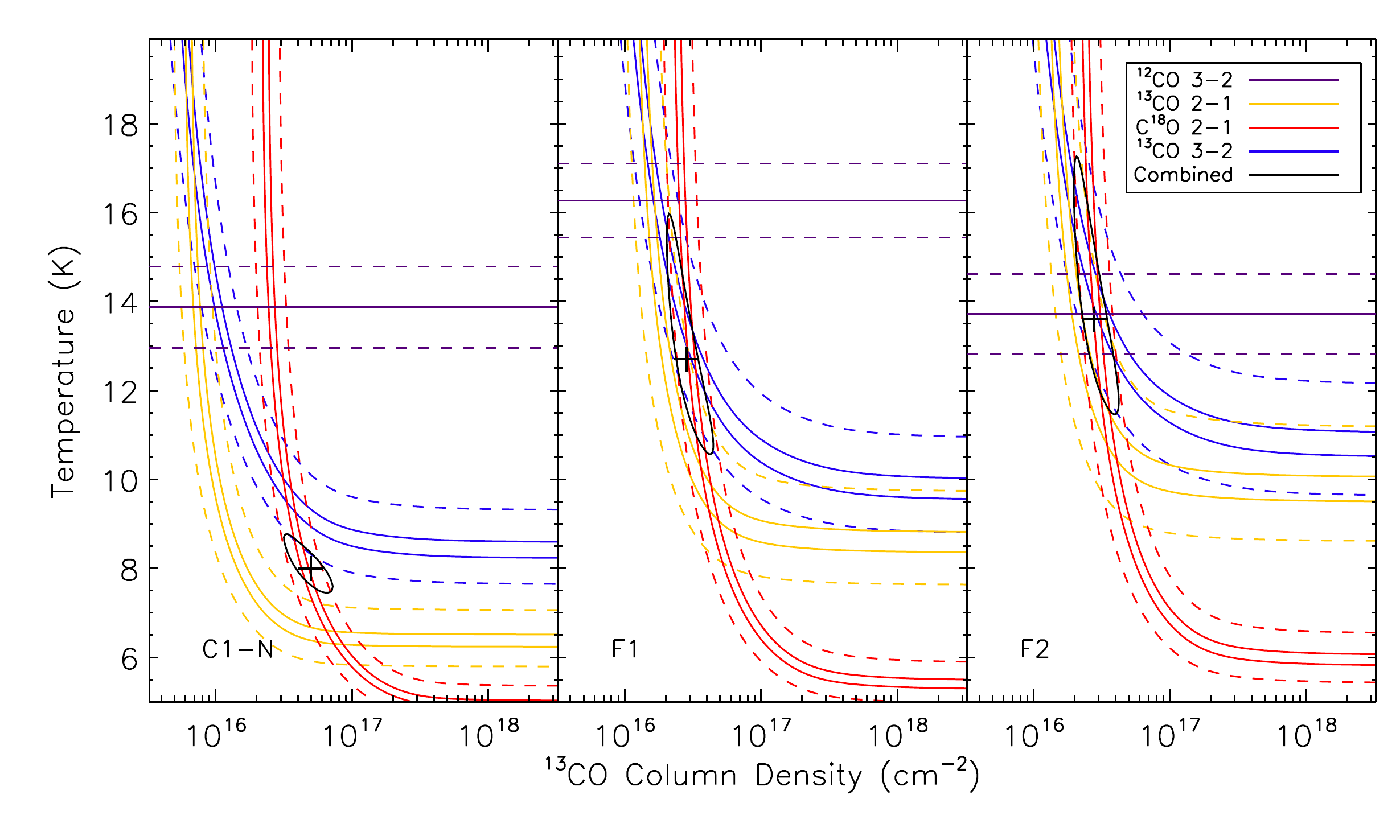}
   \caption{Best fitting gas kinetic temperature and $^{13}$CO column density ranges for the observed integrated intensities towards C1-N (left), F1 (center), and F2 (right) based on RADEX modelling. The solid yellow, red, and blue lines show the models where the integrated intensities of the $^{13}$CO J = 2 $\rightarrow$ 1, C$^{18}$O J = 2 $\rightarrow$ 1, and $^{13}$CO J = 3 $\rightarrow$ 2, respectively, were within 5\% of the observed values. The dashed yellow, red, and blue lines show where the models were within 25\% of the observed integrated intensities of the above three lines individually. The solid black ellipse shows the 1-sigma uncertainty range for models that match all three lines, found by the locus of models with a $\chi^2$ value within 1 of the best fitting model. The black cross indicates the best fitting model. The solid horizontal purple line gives the $^{12}$CO  J = 3 $\rightarrow$ 2 derived temperature while the dashed purple lines give the uncertainty on this temperature.}
   \label{fig:radexcores}
\end{figure*}

Figure \ref{fig:radexcores} shows the best fitting gas kinetic temperature and $^{13}$CO column density ranges for the observed integrated intensities towards C1-N, F1, and F2. The temperature and column density are partially degenerate, with larger column densities capable of producing the same emission if the temperature is also lowered. The combination of the three lines, however, provides a loose constraint on the kinetic temperature and column density, and the best fitting values are listed in Table \ref{table:tex}. 

To estimate an uncertainty in the gas kinetic temperature derived from the RADEX model fitting, we take the difference between the best fitting model and the model with the same column density but the largest temperature which produces a $\chi^2$ value no greater than 1 larger than for the best fitting model. A similar approach is used for the uncertainty in the $^{13}$CO column density. 
 
The RADEX models also calculate the excitation temperatures of the different lines, based upon the input gas conditions. For the $^{13}$CO J = 3 $\rightarrow$ 2 line, the excitation temperatures in the best fitting RADEX model are 0.4, 1.6 and 2.4 K lower than the gas kinetic temperature, towards C1-N, F1, and F2, respectively. 

\begin{table*}
\begin{minipage}{\textwidth}
\caption{Kinetic Temperatures}
\begin{center}
\begin{tabular}{cccccc}
\hline
\hline
Core & $^{12}$CO T & RADEX T & $\Delta$T & RADEX $N_{13}$ & RADEX Depletion\\
 & (K) & (K) & (K) & (10$^{16}$ cm$^{-2}$) & \\
(1) & (2) & (3) & (4) & (5) & (6) \\
\hline
C1-N & 13.9 (0.9) & 8.0 (0.3) & 5.9 & 4.9 (1.3) & 4.9 (1.0)  \\
F1 & 16.3 (0.8) & 12.7 (1.3) & 3.6 & 2.9 (0.6) & 2.8 (0.5) \\
F2 & 13.7 (0.9) & 13.6 (1.6) & 0.1 & 2.8 (0.6) & 4.3 (0.8) \\
\hline
\end{tabular}
\tablefoot{Column 1 gives the name of the core observed. Columns 2 and 3 give the kinetic temperatures of the gas derived from the $^{12}$CO J = 3 $\rightarrow$ 2 data and the RADEX models constrained by the rarer isotopologue lines, respectively, while Col.~4 gives the difference between the temperatures. The $^{13}$CO column density, $N_{13}$ from the best fitting RADEX model is given in Col.~5 while the corresponding CO depletion factor is given in Col.~6. The values in parentheses are the uncertainties.}
\label{table:tex}
\end{center}
\end{minipage}
\end{table*}

Radex models can simultaneously explain the $^{13}$CO and C$^{18}$O data for F1 and F2. For F2, the gas kinetic temperature derived is similar to that derived from the $^{12}$CO J = 3 $\rightarrow$ 2 line, but for F1, the less abundant isotopologues of CO require a gas temperature cooler by about 4 K. If the multiple component structure of the F1 lines were due to self-absorption, this would lower the observed integrated intensities and produce best fits with slightly lower temperatures. We consider it, however, unlikely that the full 4 K discrepancy is due to self-absorption, as the C$^{18}$O line should be optically thin, yet shows two very clear velocity components. 

For C1-N, there are no RADEX models that simultaneously provide a good prediction for all of the observed $^{13}$CO and C$^{18}$O intensities. In particular, the models do not produce as low of a $^{13}$CO J = 2 $\rightarrow$ 1 integrated intensity. This discrepancy, however, may be due to the loss of some $^{13}$CO J = 2 $\rightarrow$ 1 flux due to contamination from the reference position or from self-absorption. The 8.0 K temperature derived for C1-N from the RADEX models may thus be a slight underestimate of the temperature, although Fig.~\ref{fig:radexcores} shows that the best fitting models for the $^{13}$CO J = 3 $\rightarrow$ 2 and C$^{18}$O J = 2 $\rightarrow$ 1 lines still have gas temperatures a few Kelvin less than indicated by the $^{12}$CO data. 

These lower gas temperatures cannot be due to our adoption of gas densities lower than that in the centers of the F1 and C1-N cores, as underestimating the gas density would lead to higher derived gas temperatures. Similarly underestimating the gas density would lead to overestimating the CO column density and underestimating the CO depletion value in Sect.~\ref{depletion}.

These results suggests that at least towards F1 and C1-N, the $^{13}$CO and C$^{18}$O emission is preferentially coming from cooler gas than the $^{12}$CO J = 3 $\rightarrow$ 2 emission. Using the $^{12}$CO derived gas kinetic temperature would slightly overestimate the temperature of the gas responsible for the rarer isotopologue emission. Furthermore, the RADEX fits show that assuming the gas kinetic temperature is equal to the excitation temperature of the $^{13}$CO J = 3 $\rightarrow$ 2 line can lead to overestimates of the excitation temperature by a few Kelvin. 

\section{CO DEPLETION}
\label{depletion}

	In cold, dense conditions, CO depletes rapidly from the gas phase by freezing onto dust grains. In low mass star-forming regions, depletion factors of 5 to 15 are common (e.g., \citealt{Caselli99, Bacmann02, Crapsi05}), but the level of depletion in IRDCs is significantly less well understood. \citet{Hernandez11Tan} previously used the $^{13}$CO J = 1 $\rightarrow$ 0 transition to estimate the CO depletion factor for IRDC F. They found no significant CO freeze out, with the column density from mid-IR extinction mapping agreeing with that from the CO to within a factor of two. However, \citet{Hernandez11Tan} used a constant 20 K excitation temperature estimate for the entire cloud, which is larger than the temperatures we find in Sect.~\ref{temperature}. Using the C$^{18}$O J = 1 $\rightarrow$ 0 and 2 $\rightarrow$ 1 lines, Barnes et al.~(in prep.) find moderate depletion within IRDC F, with depletion factors up to 6. Using the SMA, \citet{Zhang09} estimate that the CO depletion may be as high as 10$^2$ to 10$^3$ towards the centers of presumably star-forming hot cores to the northeast of our surveyed area in IRDC C, although \citet{Pillai07} find a depletion factor of only two for the this region based on larger beam IRAM 30 m observations. 

\citet{Butler12} present extinction based mass surface density maps of IRDCs C and F that are believed to be accurate for mass surface densities between 0.01 g cm$^{-2}$ and 0.5 g cm$^{-2}$. These maps are based upon {\it Spitzer} data with a resolution of 2 arcsec. We smooth the \citet{Butler12} maps to match the 15 arcsec HPBW of the JCMT and regrid them to match the grid of the JCMT observations. To calculate CO depletions, only pixels where the \citet{Butler12} mass surface density is above 0.01 g cm$^{-2}$ are used, although there are only a handful of pixels at the very edges of these maps that have lower mass surface densities. These {\it Spitzer} derived maps may be slightly less accurate at very low mass surface densities compared to extinction maps based on a combination of mid-infrared and near-infrared data (e.g., \citealt{Kainulainen13Tan}), but should have comparable accuracy for mass surface densities above 0.05 g cm$^{-2}$, as appropriate for most of the observed regions and for where CO depletion is expected to be most prominent. 

\subsection{Depletion Towards the C1-N, F1, and F2 Cores}

Towards the C1-N, F1, and F2 cores, the $^{13}$CO column density was already estimated in Sect.~\ref{temperature}. These $^{13}$CO column densities can be converted to $^{12}$CO columns given the assumed isotopologue ratios of 43 and 51 for IRDC C and F, respectively (see also Sect.~\ref{temperature}). Based upon the abundance gradients in the Galactic Disk published by \citet{Wilson92} and the solar neighbourhood abundance of $^{12}$CO from \citet{Frerking82}, \citet{Fontani06} find that the ratio of CO to H$_2$ molecules is given by
\begin{equation}
^{12}\mbox{CO} / \mbox{H}_2 = 9.5 \times 10^{-5} \exp\left(1.105 - 0.13 D_\text{GC}(\mbox{kpc})\right),
\end{equation}
where $D_\text{GC}$ is the Galactocentric radius of the cloud. \citet{Miettinen11} and \citet{Fontani12} also uses this relation\footnote{The original relation given by \citet{Frerking82} uses a coefficient of 8.5, rather than 9.5, and this lower coefficient is used by \citet{Fontani12}. The 9.5 coefficient is based upon the updated C$^{18}$O to $^{12}$CO ratio determined by \citet{Wilson92}.}. IRDCs C and F are at Galactocentric radii of 4.7 kpc and 5.8 kpc, such that the $^{12}$CO abundance should be 1.6 and $1.3 \times 10^{-4}$, respectively, within these clouds. This is consistent with the canonical values of 1 and $2 \times 10^{-4}$ found towards Taurus and Orion \citep{Frerking82,Lacy94}. For a mean mass per H$_2$ molecule of $4.6 \times 10^{-24}$ g, or about 2.77 amu \citep{Kaufman99}, surface densities in g cm$^{-3}$ can then be derived to compare to those from \citet{Butler12}. The CO depletion factors for these three cores, given as the ratio of the extinction derived mass surface density to the CO derived mass surface density, are given in Table \ref{table:tex}. The uncertainties in these depletion factors are found from propagating the uncertainties in the $^{13}$CO column densities. The random uncertainties in the \citet{Butler12} column densities are small compared to the large uncertainty in the assumed excitation temperature for the $^{13}$CO. Systematic uncertainties in either the extinction or CO data, however, could easily be at the level of a factor of two. 

We find that CO is frozen out towards the C1-N, F1, and F2 cores, with depletion factors between 2 and 5. The maximum depletion at the centers of these cores could be much larger than these values, as our CO observations are likely tracing much of the envelope along the line of sight towards these cores. 

\subsection{Depletion Throughout IRDCs C and F}

To estimate the CO depletion factor throughout the JCMT observed regions of IRDCs C and F, the $^{12}$CO J = 3 $\rightarrow$ 2 transition cannot be used to derive column densities as it is highly optically thick. Conversely, the C$^{18}$O J = 3 $\rightarrow$ 2 is detected in too few locations, and even then only detected with very low signal to noise, to provide reliable column density estimates. As such, we use the $^{13}$CO J = 3 $\rightarrow$ 2 transition to derive CO based column densities throughout IRDCs C and F.

For each detected component in the $^{13}$CO J = 3 $\rightarrow$ 2 spectra, a grid of RADEX models are run with FWHM set to that of the component. As in Sect.~\ref{temperature}, the H$_2$ density is set to $5 \times 10^4$ cm$^{-3}$ for IRDC C and $2 \times 10^{4}$ cm$^{-3}$ for IRDC F \citep{Butler09}. Two thousand models spaced logarithmically over 6 orders of magnitude in column density are run for each component and the model that produces an integrated intensity closest to that observed is selected as the best fitting model. The $^{13}$CO column density towards a given pointing is taken to be the sum of the column densities derived for each component towards that point. The gas kinetic temperature of the models is initially set to be equal to the kinetic temperature derived from the $^{12}$CO J = 3 $\rightarrow$ 2 transition for each pointing location. Such a temperature should be a reasonable upper limit for the kinetic temperature of the gas, which should produce upper limits for the depletion factor. To estimate a lower limit for the depletion, the models are rerun with the kinetic temperature set to 4 K less than the $^{12}$CO derived temperature, similar to that derived towards the F1 and C1-N cores. The RADEX code self-consistently calculates the $^{13}$CO J = 3 $\rightarrow$ 2 excitation temperature, taking into account possible sub-thermal excitation.

\begin{figure*}
   \centering
   \begin{subfigure}[b]{0.49\textwidth}
     \centering
      \includegraphics[width=3.5in]{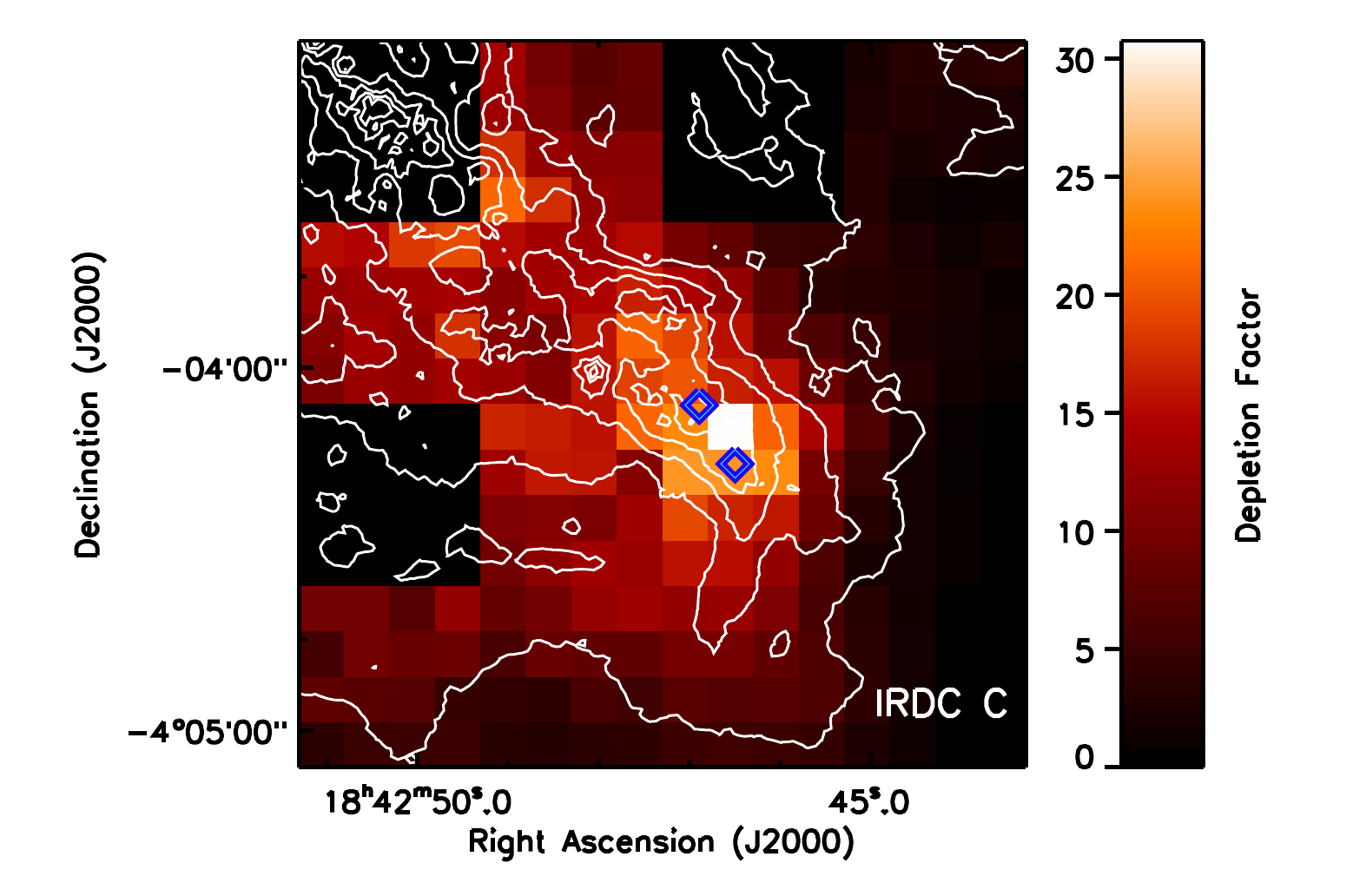}
   \end{subfigure}
   \begin{subfigure}[b]{0.49\textwidth}
      \centering
      \includegraphics[width=3.5in]{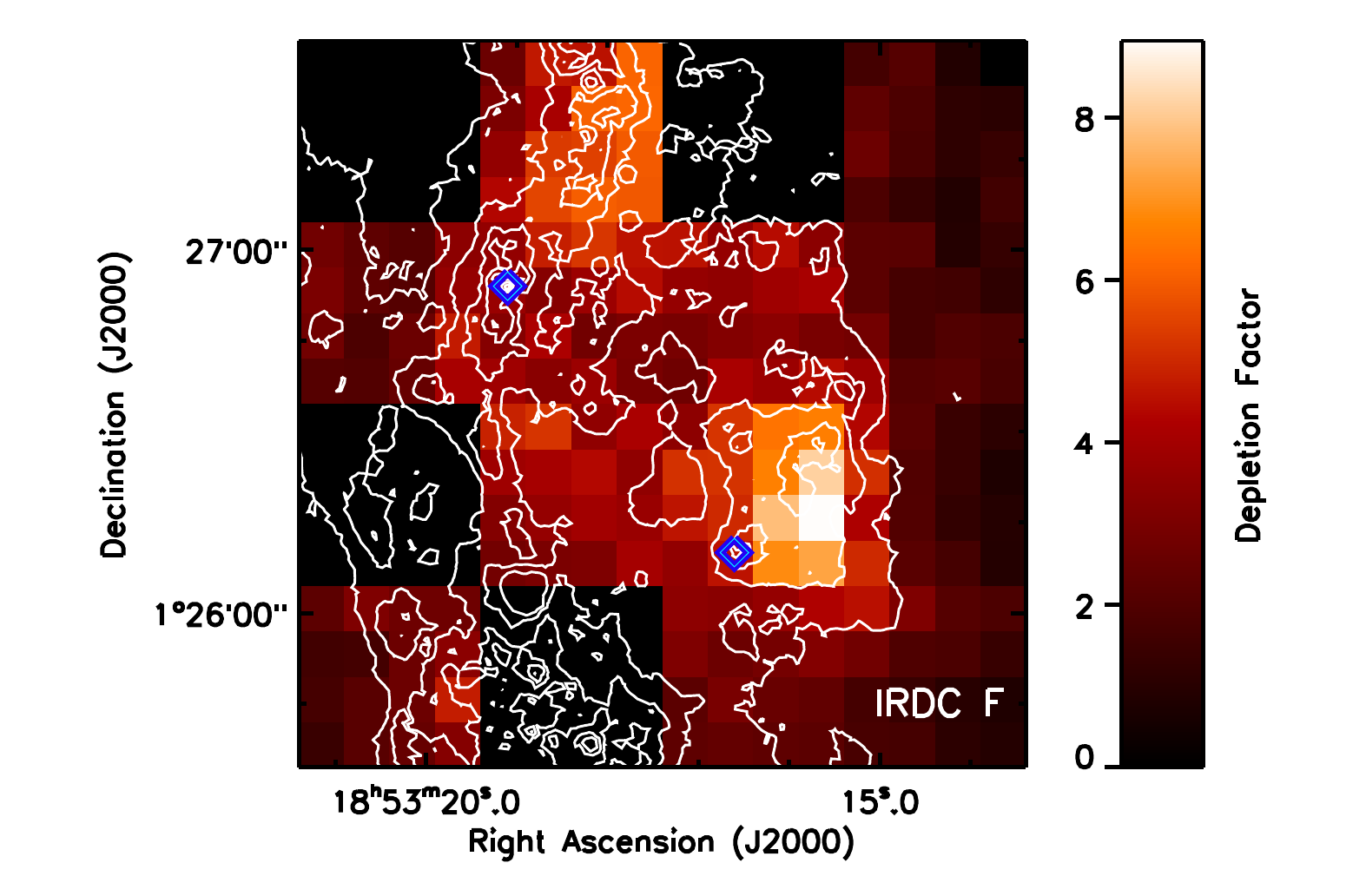}
   \end{subfigure}
   \caption{CO depletion factor towards IRDC C ({\it left}) and IRDC F ({\it right}) derived under the assumption that the kinetic temperature for the $^{13}$CO is the same as the $^{12}$CO. The contours give the mass surface density derived by \citet{Butler12} and are the same as in Fig.~\ref{fig:sigmaradec}. The blue diamonds give the central locations of the C1-N, C1-S, F1, and F2 cores.}
   \label{fig:depletiontoff0}
\end{figure*}

\begin{figure*}
   \centering
   \begin{subfigure}[b]{0.49\textwidth}
     \centering
      \includegraphics[width=3.5in]{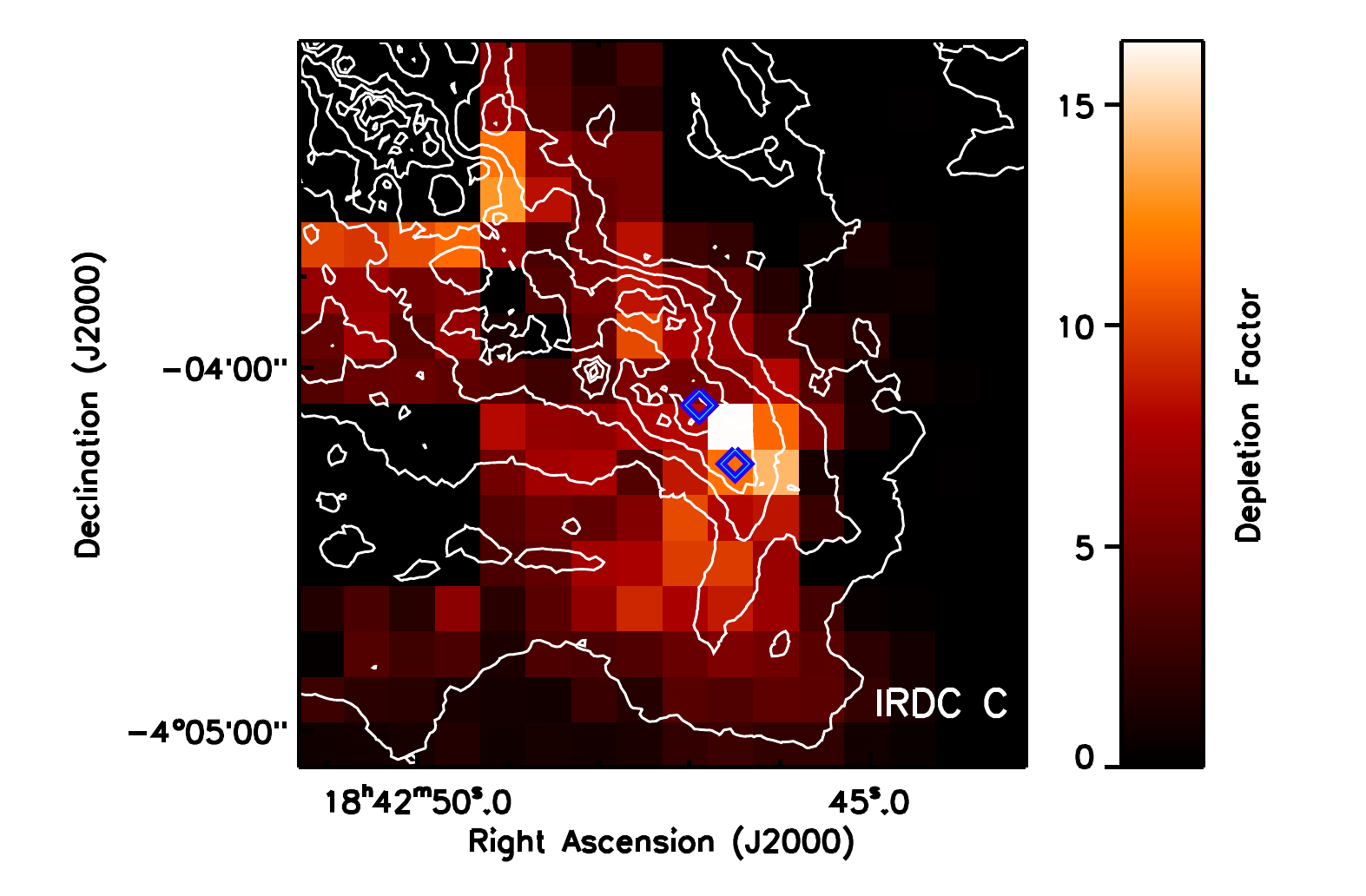}
   \end{subfigure}
   \begin{subfigure}[b]{0.49\textwidth}
      \centering
      \includegraphics[width=3.5in]{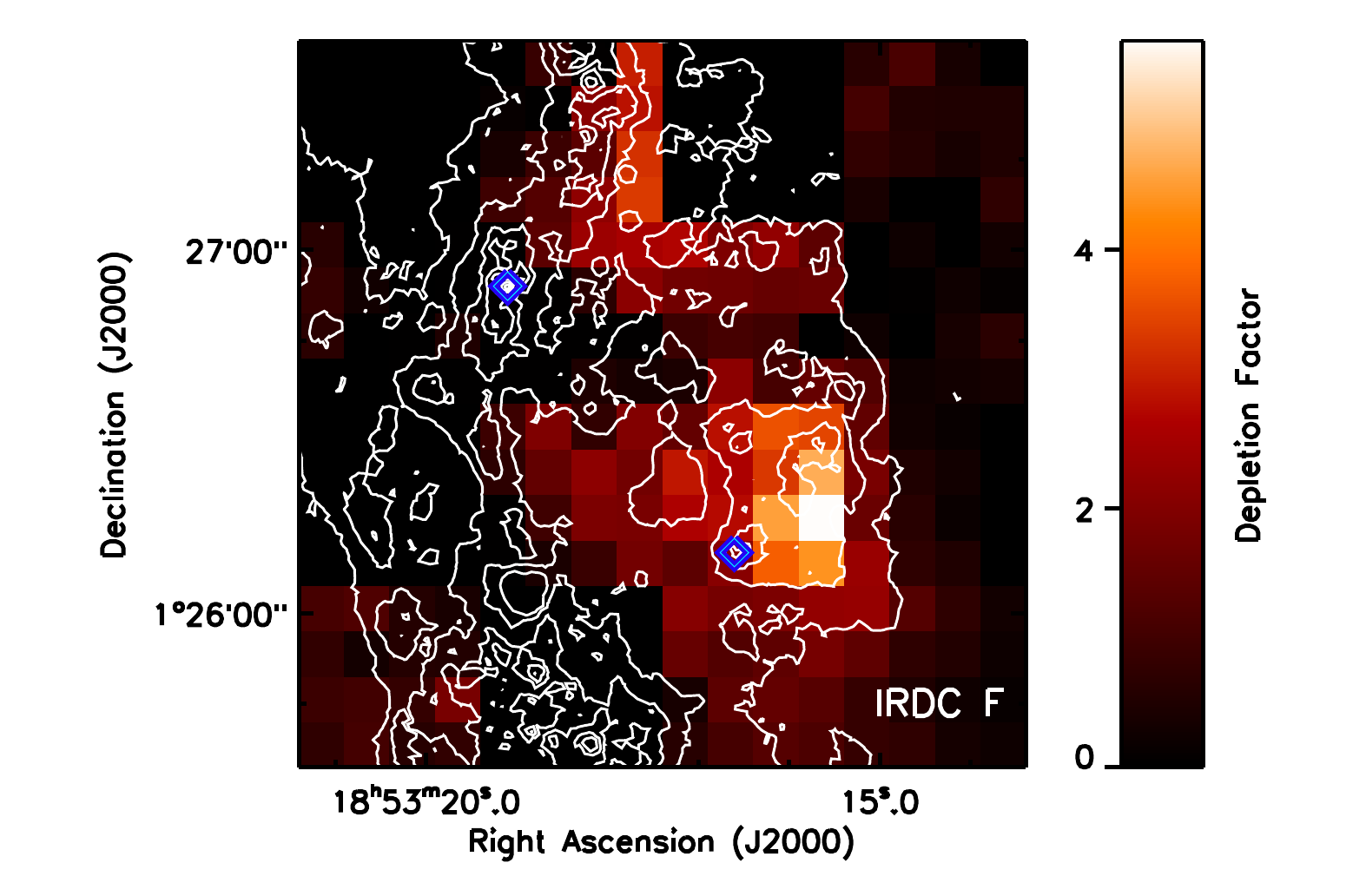}
   \end{subfigure}
   \caption{CO depletion factor towards IRDC C ({\it left}) and IRDC F ({\it right}) derived under the assumption that the kinetic temperature for the $^{13}$CO is 4 K less than for the $^{12}$CO. The contours give the mass surface density derived by \citet{Butler12} and are the same as in Fig.~\ref{fig:sigmaradec}. The blue diamonds give the central locations of the C1-N, C1-S, F1, and F2 cores.}
   \label{fig:depletiontoff4}
\end{figure*}

\begin{figure*}
   \centering
   \begin{subfigure}[b]{0.49\textwidth}
     \centering
      \includegraphics[width=3.5in]{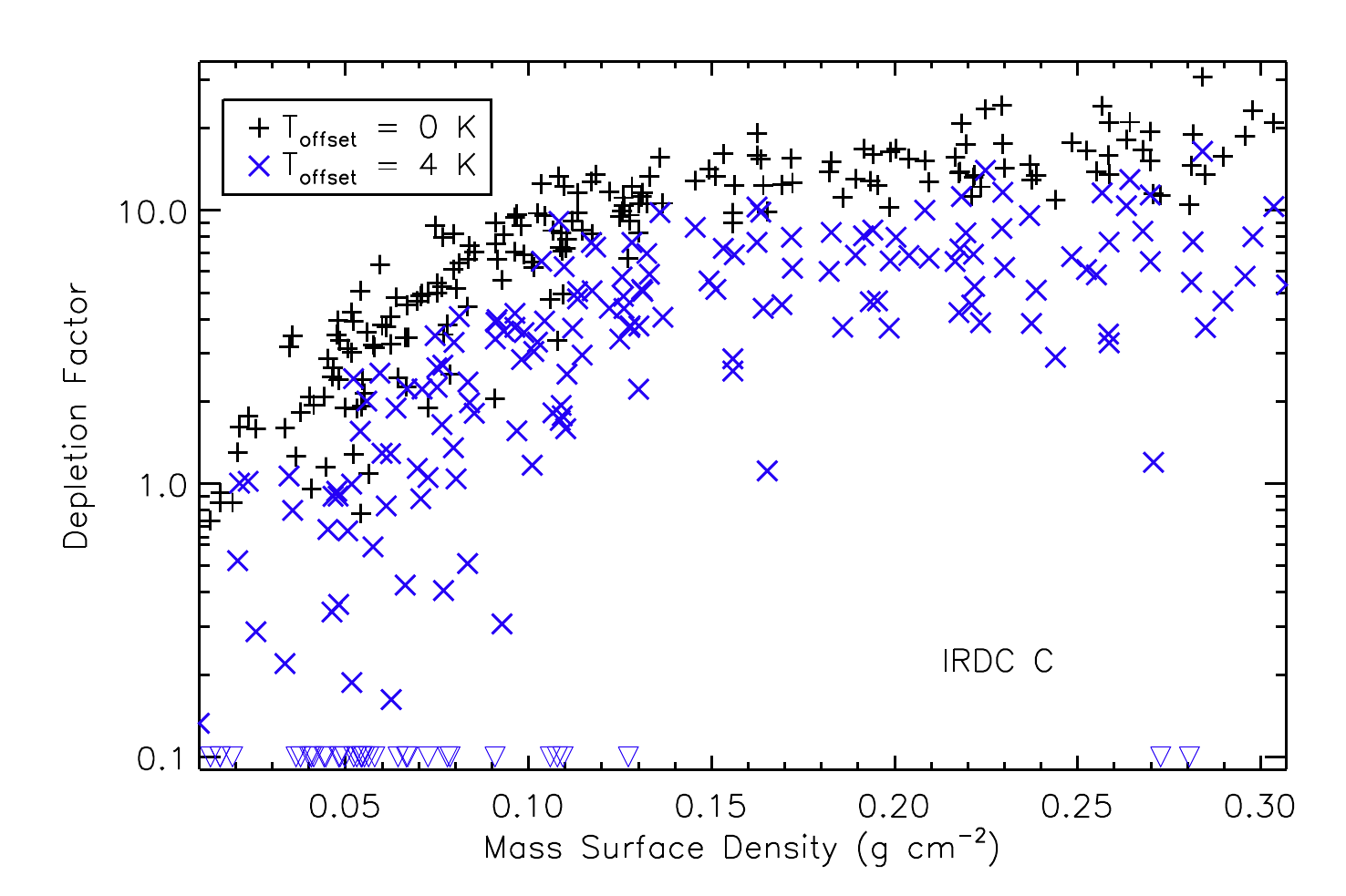}
   \end{subfigure}
   \begin{subfigure}[b]{0.49\textwidth}
      \centering
      \includegraphics[width=3.5in]{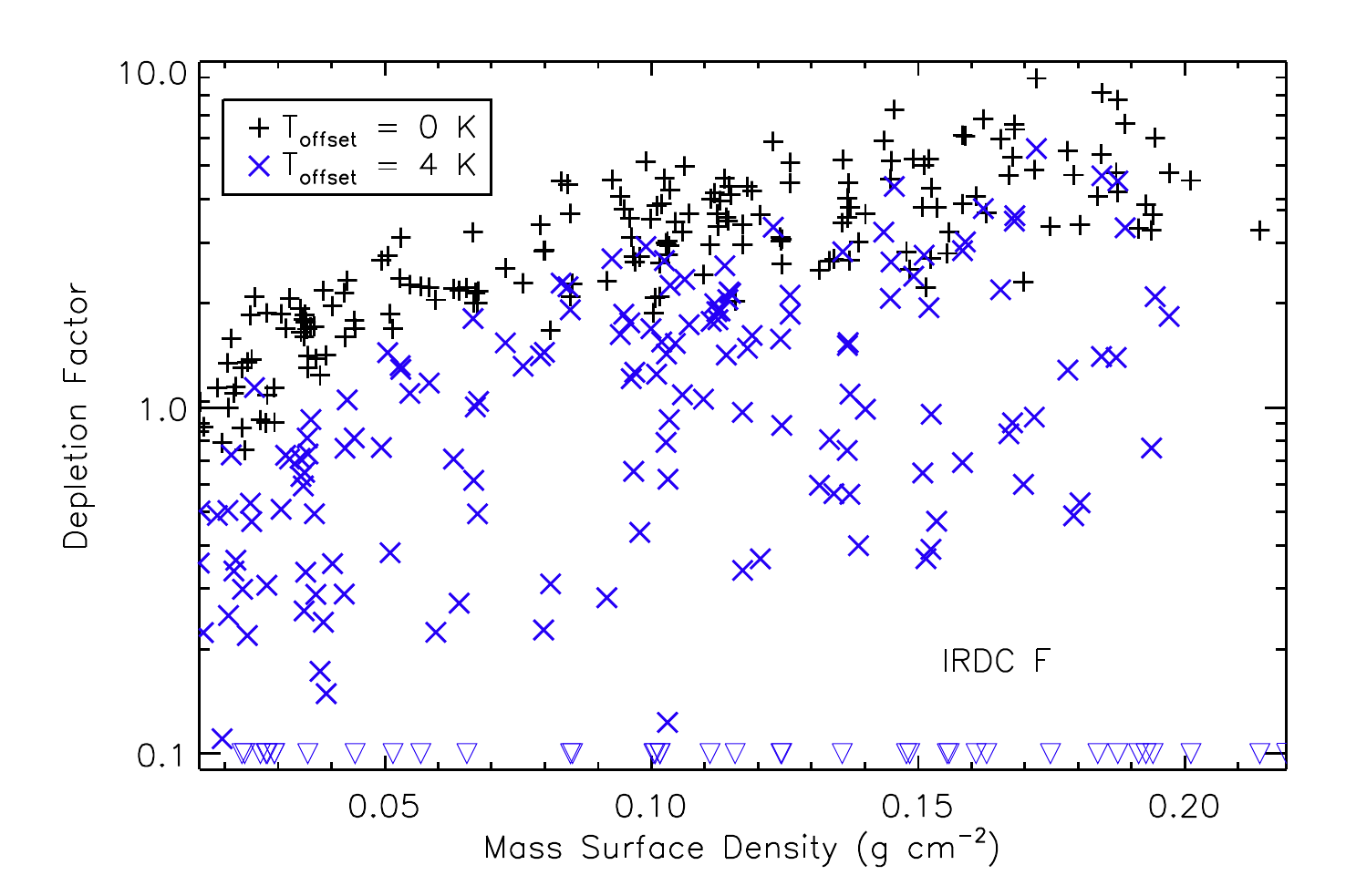}
   \end{subfigure}
   \caption{CO depletion factors as a function of extinction based mass surface density for IRDC C ({\it left}) and IRDC F ({\it right}). The black crosses show the depletion factors derived assuming no temperature offset between the $^{13}$CO and $^{12}$CO while the blue x's are derived assuming that $^{13}$CO traces gas 4 K colder than the $^{12}$CO. Triangles indicate that the depletion factor was lower than 0.1. Note the larger depletion factors derived for IRDC C.}
   \label{fig:dvscol}
\end{figure*}

Figures \ref{fig:depletiontoff0} and \ref{fig:depletiontoff4} show the depletion factors derived assuming no temperature offset between the $^{13}$CO and $^{12}$CO and assuming a 4 K difference between the $^{13}$CO and $^{12}$CO, respectively. Figure \ref{fig:dvscol} shows how the derived depletion factors vary as a function of the extinction derived mass surface density. We find that there is a substantial level of CO depletion throughout both IRDC C and F. The depletion factor clearly increases towards the C1 and F1 clumps, although the F2 clump does not show significantly increased depletion factors relative to its surroundings. In the F1 clump, the peak depletion occurs to the west of the F1 core, which is why the F1 core depletion listed in Table \ref{table:tex} is even lower than that for the F2 core. 

\

If CO is not depleted at all, the depletion factor should be one, while the freezing out of CO should produce $f_\text{d}$ values greater than one. Any values of $f_\text{d}$ less than one could be interpreted as indicating CO gas abundances larger than canonically expected, but are more likely due to overestimates in the mass surface density from the CO data, possibly due to underestimated temperatures. For the depletion factors derived from assuming no temperature offset, the depletion factors are almost always greater than one, with $f_\text{d}$ approaching one towards the outskirts of the observed regions. For an offset of 4 K, the depletion factor, particularly towards lower column densities in IRDC C, regularly drops below one. In particular, note that towards the F2 core, where the IRAM data ratios indicate that the $^{13}$CO gas temperature is roughly equal to that of the $^{12}$CO, a depletion factor less than one is produced when an offset temperature of 4 K is used. This suggests that for large portions of IRDC C and F, the $^{13}$CO J = 3 $\rightarrow$ 2 transition probes gas at similar temperatures to that of the $^{12}$CO J = 3 $\rightarrow$ 2 line. The significant uncertainty as to which temperature should be used means that the derived depletion factors should be considered to be only valid to within factors of a few. Since the depletion factor in the outskirts of the maps approaches one under the assumption of equal excitation temperature, the depletion factors in Fig.~\ref{fig:depletiontoff0} should be considered equivalent to the relative depletion factors derived by \citet{Hernandez11Caselli, Hernandez12} by setting the depletion factor to one at the outskirts of their observed IRDC.

Figure \ref{fig:dvscol} shows a distinct trend for the depletion factor to increase with increasing column density, as expected. The maximum depletion factor towards IRDC F lies within the range of 5 to 9, while the peak depletion factor towards IRDC C is at least 16, but not greater than 31. In general, IRDC C shows greater levels of CO depletion than throughout IRDC F, with many of the high column density sight lines towards IRDC C showing depletion values greater than 5. IRDC F seems to have a similar level of depletion as typically seen in low mass star-forming regions (depletion factors of the order of 5 to 10, \citealt{Caselli99, Crapsi05}), while IRDC C has a larger degree of CO depletion than is common in these lower mass star-forming regions. The higher level of depletion in IRDC C, however, is still consistent with the range of CO depletion derived previously for infrared dark clouds (depletion factors up to 78, \citealt{Fontani12}).

If the $^{13}$CO J = 3 $\rightarrow$ 2 spectra are affected by self-absorption, the above derived CO column densities will be slightly underestimated, such that the CO depletion factors will be slightly overestimated. See Sect.~\ref{kinematics}, however, for discussion regarding why the multiple detected components may be due to separate structures along the line of sight, rather than due to self-absorption. Uncertainties in the adopted $^{12}$CO abundance, isotopologue ratios, and mean mass per hydrogen molecule may produce a systematic uncertainty up to a factor of two, but should be smaller than the uncertainty in the depletion factor induced by the uncertain kinetic temperature of the gas probed by the $^{13}$CO. Similarly, further uncertainties from the gas-to-dust ratio and opacity per unit mass value, $\kappa_{8\mu m}$, adopted by \citet{Butler12} should be minor in comparison.

The choice of using a constant density throughout the map will generally overestimate the density in the periphery of the map and underestimate the density towards the centers of cores. At higher densities, the same column of material will produce more emission, or, interpreted the other way around, the same amount of emission can be explained by a lower column density of gas at a higher density. Thus, underestimating the density towards the core centers would underestimate the depletion, while overestimating the density in the periphery of the maps would overestimate the depletion. 

Overall, we find that CO is depleting from the gas phase within IRDC C and F, to an extent potentially larger than seen in lower mass star-forming regions. Such CO depletion thus appears to be a common part in the evolution of the gas phase chemistry within IRDCs. 

\section{KINEMATICS}
\label{kinematics}

\subsection{Global Contraction or Expansion}

	The $^{12}$CO J = 3 $\rightarrow$ 2 spectra are highly optically thick and much of the line structure observed is likely due to self-absorption. For sources with excitation temperature gradients, with the excitation temperature increasing towards the center, infall and outflow motions will produce asymmetric line profiles in optically thick lines. Infalling sources will produce brighter blue peaks, while outflows will produce brighter red peaks (e.g., \citealt{Mardones97}). Optically thin lines would then be expected to appear between the two components, if the asymmetry is due to self-absorption.
	
\begin{figure}
   \centering
   \includegraphics[width = 3.5 in]{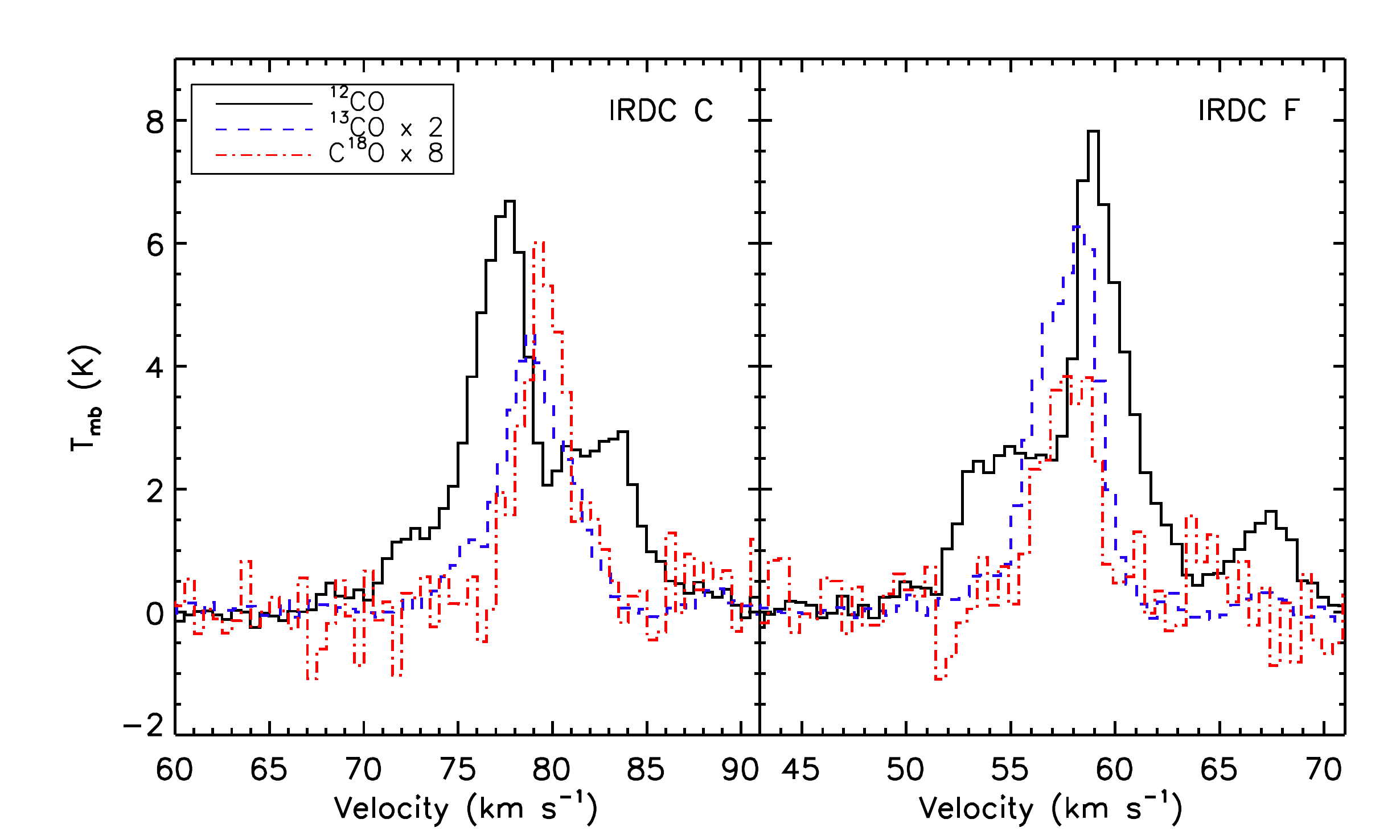}
   \caption{Spatially averaged spectra of the $^{12}$CO (black), $^{13}$CO (blue, dashed), and C$^{18}$O (red, dash-dotted) J = 3 $\rightarrow$ 2 lines over IRDC C (left) and IRDC F (right). The $^{13}$CO data have been scaled up by a factor of two while the C$^{18}$O data have been scaled up by a factor of 8.}
    \label{fig:overlay}
\end{figure}

	Figure \ref{fig:overlay} shows the spatially averaged $^{12}$CO, $^{13}$CO, and C$^{18}$O J = 3 $\rightarrow$ 2 spectra for IRDCs C and F. The IRDC C average $^{12}$CO spectrum clearly shows a larger blue peak, with the $^{13}$CO and C$^{18}$O lines centered between the red and blue $^{12}$CO peaks, in the classical picture of an infalling source. This blue asymmetry is seen in numerous individual spectra across IRDC C, including towards C1-S and C1-N. It is, however, unlikely that this asymmetry probes local collapse of the C1-S and C1-N cores since the asymmetry is seen so widely across the map. Rather, this may indicate a more global collapse of this region of the IRDC. 
	
	In IRDC F, the average spectrum shows a less pronounced asymmetry, with the $^{13}$CO and C$^{18}$O lines centered only slightly to lower velocities than the main $^{12}$CO peak, with a smaller blue shoulder also existing in the $^{12}$CO line. As shown in Appendix \ref{appendix:jcmtfits}, when two components are clearly visible in the $^{12}$CO line, the red component is almost always the brighter of the two in IRDC F. One may thus think that the observed region of IRDC F is globally expanding. Such a simple interpretation of line asymmetries for IRDCs C and F, however, may not fully hold in such complicated, large regions, especially given the presence of multiple different gas components, as described below. 

\subsection{Multiple Components}

	The C$^{18}$O data has relatively low signal to noise and will not be further examined for kinematic information. The $^{13}$CO J = 3 $\rightarrow$ 2 line, on the other hand, is only moderately optically thick ($\tau \sim 3$) and is strongly detected, such that it provides a reasonable probe of potential substructures within the observed IRDCs. 
	
\begin{figure}
   \centering
   \begin{subfigure}[b]{0.5\textwidth}
      \centering
   \includegraphics[width = 3.5 in]{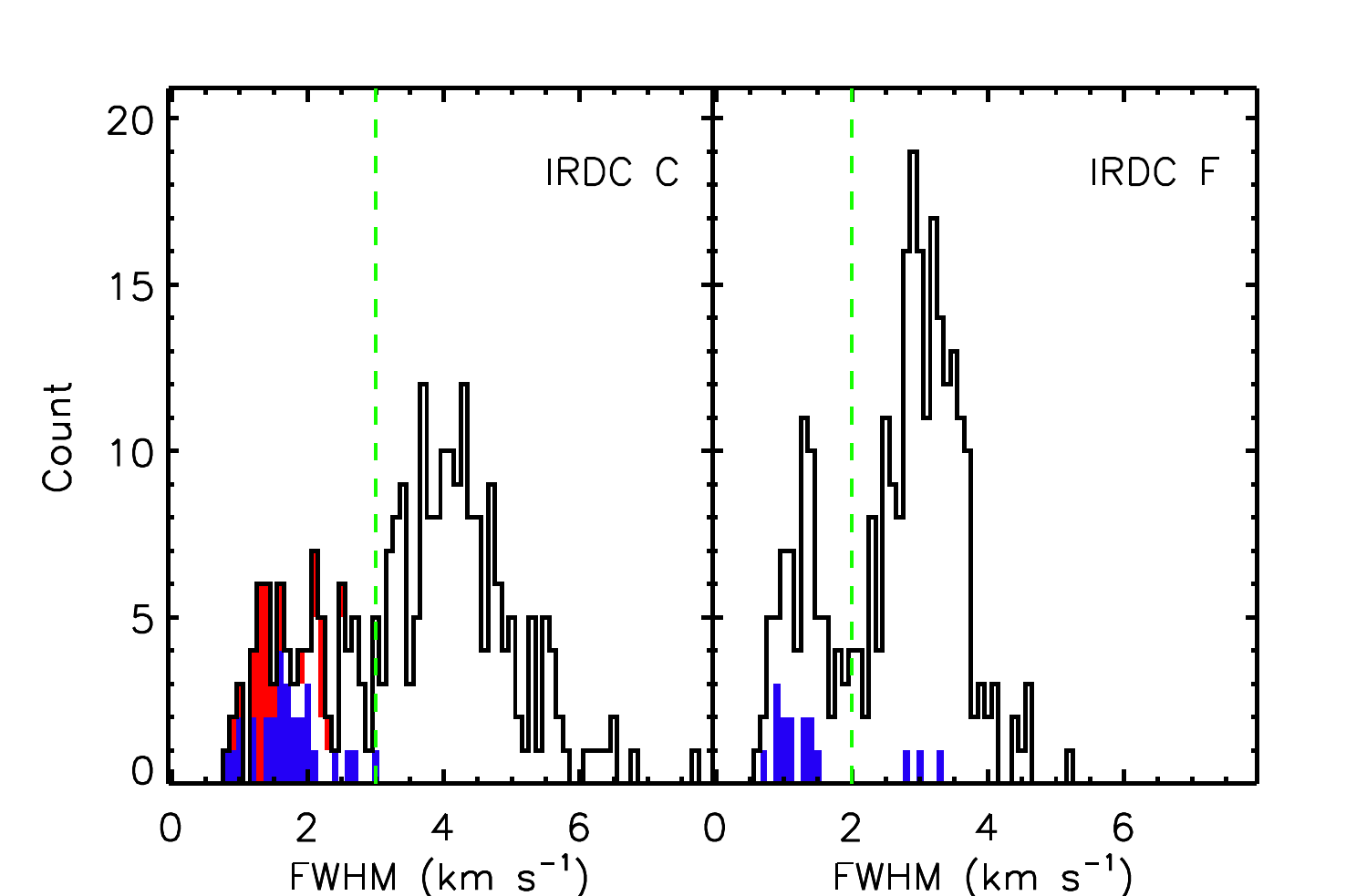}
   \end{subfigure}
        \begin{subfigure}[b]{0.5\textwidth}
           \centering
     \includegraphics[width= 3.5 in]{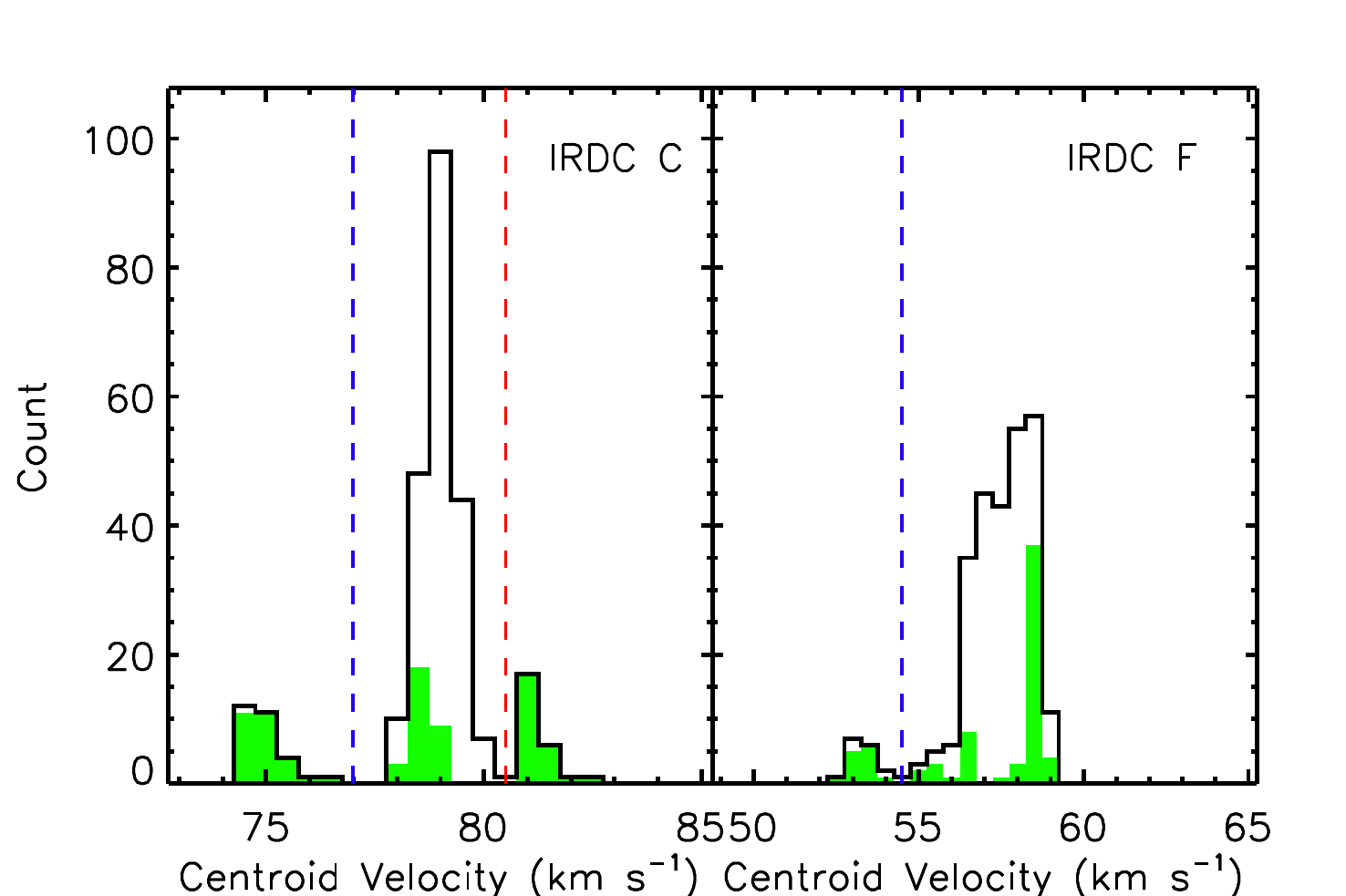}
     \end{subfigure}
   \caption{Histograms of the occurrence of different FWHM (top) and centroid velocity (bottom) of the different detected line components of $^{13}$CO J = 3 $\rightarrow$ 2. The left column is for IRDC C and the right column for IRDC F. In the FWHM histogram for IRDC C, components with centroid velocities greater than 80.5 km s$^{-1}$ are shown in light red, velocities less than 77 km s$^{-1}$ in dark blue and intermediate velocities in white. For IRDC F, in the FWHM histogram, components with centroid velocities less than 54.5 km s$^{-1}$ are shown in blue. These threshold velocities are denoted as dashed vertical lines in the bottom panels in the corresponding colours. For the VLSR histograms, components with a FWHM less than 3 km s$^{-1}$ in IRDC C and less than 2 km s$^{-1}$ in IRDC F are shown in green. This dividing width is denoted as the vertical green line in the top panels.}
    \label{fig:histograms}
\end{figure}

\begin{figure*}
\centering \includemovie[
     3Dviews2=test.tex,
        toolbar, 
        label=c13copeak.u3d,
     text={\includegraphics[width=150mm]{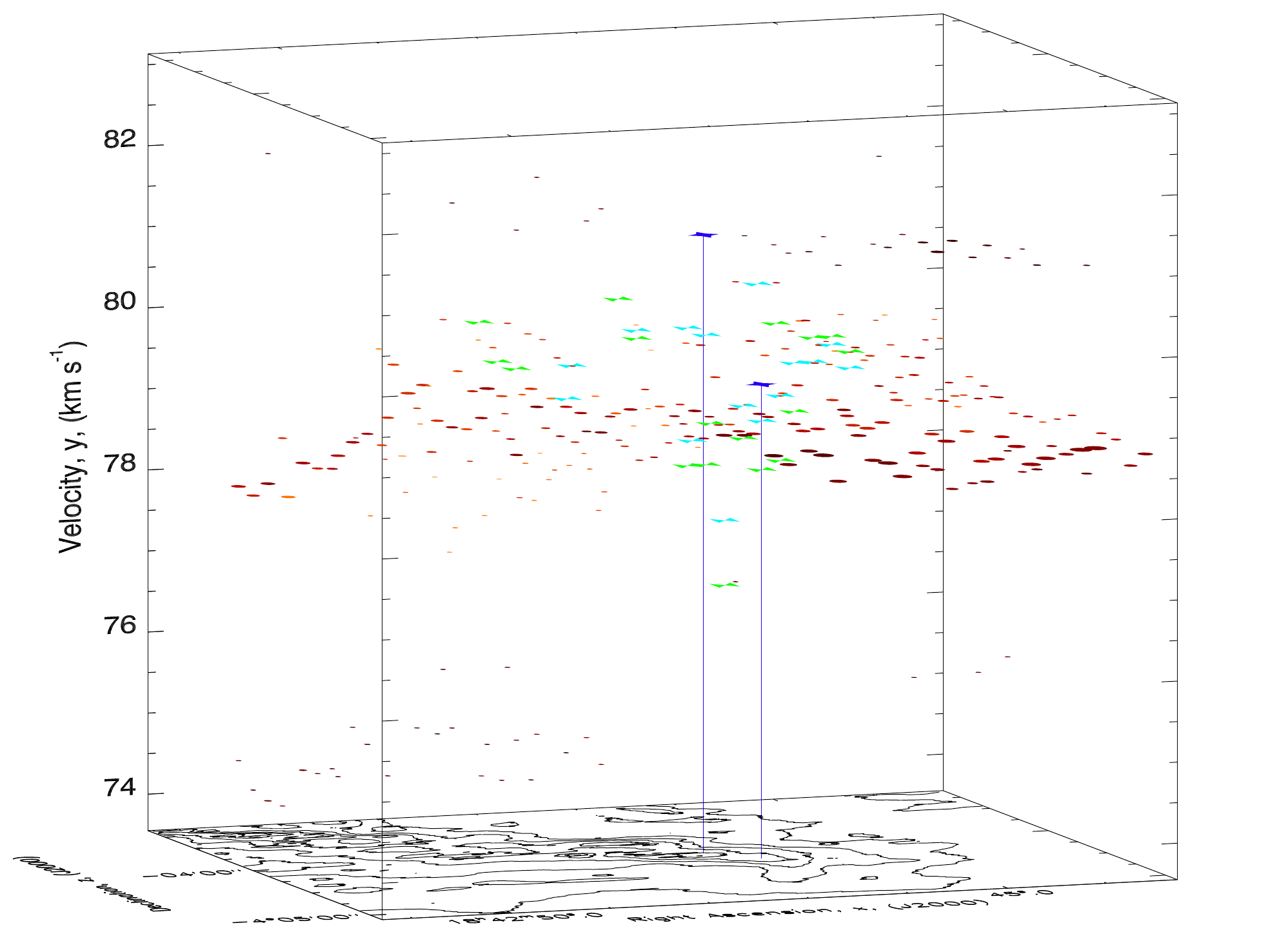}},
        3Dlights=CAD,
]{}{}{c13copeak.u3d}
\caption{Right ascension-declination-velocity diagram for the detected lines towards IRDC C. The circular points shaded from black to light orange are the $^{13}$CO components detected by the JCMT. For these JCMT points, the size of the point is scaled to the peak intensity, with larger points corresponding to larger intensities, while the colour is scaled based upon the FWHM, with lighter colours denoting larger FWHM. The bow-tie shaped green and light blue points (cubes in the interactive view) show the $^{12}$CO J = 8 $\rightarrow$ 7 and 9 $\rightarrow$ 8 lines detected by {\it Herschel} (Paper I), while the dark blue, star shaped points (cones in the interactive view) are the N$_2$D$^+$ J = 3 $\rightarrow$ 2 detections of the C1-N and C1-S cores by \citet{Tan13}. The C1-N core is the core at the larger declination and with the larger centroid velocity. A line has been drawn connecting the N$_2$D$^+$ points to the lower surface to better illustrate the position of the cores. The contours shown on the bottom surface are the mass surface density contours from \citet{Butler12}. If the electronic version is viewed with Adobe Acrobat, the figure can be clicked on to activate an interactive, rotatable, 3D representation of the data. In this interactive view,  the X-axis is increasing right ascension, the Z-axis is increasing declination, and the Y-axis is increasing central velocity. Right clicking and selecting disable content will return to the original static, 2D representation of the data.}
\label{fig:c13coppv}
\end{figure*}

\begin{figure*}
\centering \includemovie[
     3Dviews2=test2.tex,
        toolbar, 
        label=f13copeak.u3d,
     text={\includegraphics[width=150mm]{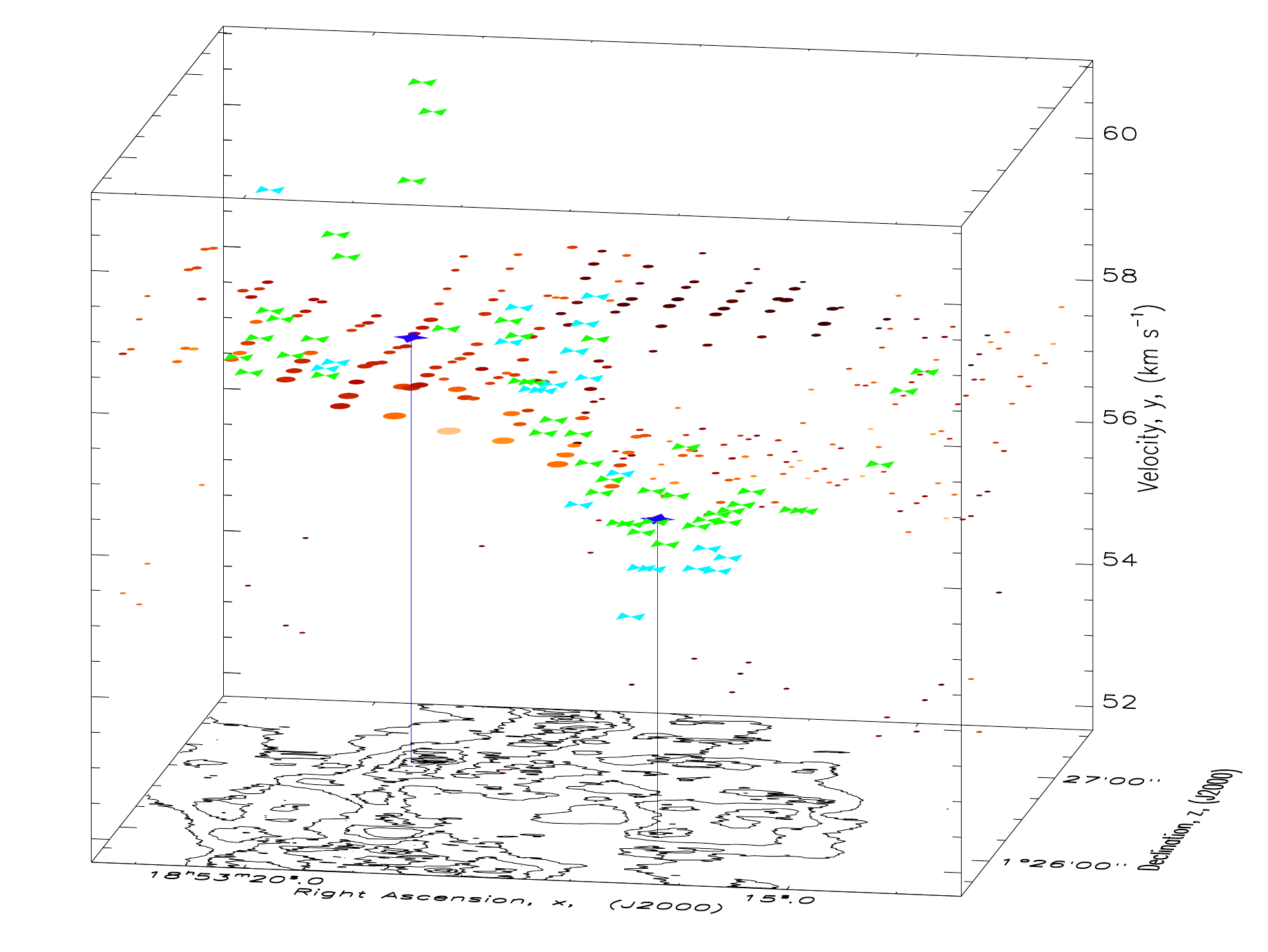}},
        3Dlights=CAD,
]{}{}{f13copeak.u3d}
   \caption{Right ascension-declination-velocity diagram for the detected lines towards IRDC F. The circular points shaded from black to light orange are the $^{13}$CO components detected by the JCMT. For these JCMT points, the size of the point is scaled to the peak intensity, with larger points corresponding to larger intensities, while the colour is scaled based upon the FWHM, with lighter colours denoting larger FWHM. The bow-tie shaped green and light blue points (cubes in the interactive view) show the $^{12}$CO J = 8 $\rightarrow$ 7 and 9 $\rightarrow$ 8 lines detected by {\it Herschel} (Paper I), while the dark blue, star shaped points (cones in the interactive view) are the N$_2$D$^+$ J = 3 $\rightarrow$ 2 detections of the F1 and F2 cores by \citet{Tan13}. The F2 core is the core at the larger declination and with the larger centroid velocity. A line has been drawn connecting the N$_2$D$^+$ points to the lower surface to better illustrate the position of the cores. The contours shown on the bottom surface are the mass surface density contours from \citet{Butler12}. If the electronic version is viewed with Adobe Acrobat, the figure can be clicked on to activate an interactive, rotatable, 3D representation of the data. In this interactive view,  the X-axis is increasing right ascension, the Z-axis is increasing declination, and the Y-axis is increasing central velocity. Right clicking and selecting disable content will return to the original static, 2D representation of the data.}
   \label{fig:f13coppv}
\end{figure*}

	Figure \ref{fig:histograms} shows histograms of the frequency of the different FWHM and central velocities of the detected line components. To supplement these histograms and to reveal any possible spatial correlations, Figs.~\ref{fig:c13coppv} and \ref{fig:f13coppv} show position-position-velocity (PPV) cubes of the $^{13}$CO data (Interactive, 3D versions of these figures can be downloaded in the online version). In these PPV plots, the JCMT data are colour coded from black to red with the colour scale set by the FWHM of the components, with larger FWHM corresponding to lighter oranges and smaller FWHM being closer to black. The size of each point is scaled to the peak intensity of the component, with the larger points being components with larger intensities. The green and light blue points show the $^{12}$CO J = 8 $\rightarrow$ 7 and 9 $\rightarrow$ 8 lines detected by {\it Herschel} (Paper I), respectively, but are not scaled in size or colour based upon their FWHM or intensity. The regions observed with Herschel are smaller than the JCMT regions. The dark blue points mark the N$_2$D$^+$ ALMA detections of the F1, F2, C1-N, and C1-S cores by \citet{Tan13}. A line has been drawn connecting the N$_2$D$^+$ points to the lower surface to better illustrate the position of the cores. The contours shown on the bottom surface are mass surface density contours \citep{Butler12}. 
		
	As seen in Fig.~\ref{fig:histograms}, while most components in IRDC C are detected at a velocity of 78 to 80 km s$^{-1}$, there are two additional groups of components centered around velocities of 75 km s$^{-1}$ and 81 km s$^{-1}$. The main velocity group tends to have FWHM of approximately 4 km s$^{-1}$, while the secondary velocity groups preferentially have smaller FWHM under 2 km s$^{-1}$. The main group of centroid velocities agree well with the 79.4 km s$^{-1}$ centroid velocity of the C1-S core, as observed in N$_2$D$^+$ J = 3 $\rightarrow$ 2 emission \citep{Tan13}. The C1-N core, however, was found to have a centroid velocity of 81.2 km s$^{-1}$ \citep{Tan13}, thereby associating the C1-N core with the weaker detections in the 81 km s$^{-1}$ group. These N$_2$D$^+$ observations suggest that multiple velocity structures exist within IRDC C, rather than the multiple velocity groups seen in the CO data being formed solely from self-absorption, and suggest that the C1-N and C1-S cores are associated with gas structures at slightly different velocities. The 75 km s$^{-1}$ group is preferentially detected in the top left corner of the map, towards the bulk of the rest of IRDC C, and may be associated with a further gas structure within IRDC C. The higher J CO detections from Herschel seem to be preferentially associated with the primary 79 km s$^{-1}$ group, but the centroid velocities of these higher J lines are not well constrained due to the low signal to noise of the 8 $\rightarrow$ 7 and 9 $\rightarrow$ 8 lines (Paper I). 
	
\citet{Wang06} detect a water maser near the C1-S core with a velocity, relative to the local standard of rest, of 59 km s$^{-1}$, which is significantly offset from the rest velocity of IRDC C, 78.6 km s$^{-1}$ \citep{Simon06}. This water maser detection is relatively weak and the water maser is not detected by the more sensitive survey of \citet{Chambers09}. While this lack of detection by \citet{Chambers09} could be due to temporal variability of the maser, as masers can dim by orders of magnitude on the timescale of years \citep{Wang12}, there is no sign of any emission at this 59 km s$^{-1}$ velocity in any of our three CO datasets, suggesting that this maser may not be real.
	
	While the centroid velocity histogram for IRDC F shows a very small velocity group at 53 km s$^{-1}$ and a broad group between 56 and 59 km s$^{-1}$, Fig.~\ref{fig:f13coppv} shows that this broad velocity group breaks into two clearly distinct gas structures towards the F1 clump. One gas structure is characterized by thin lines, which primarily account for the 1 km s$^{-1}$ FWHM peak in Fig.~\ref{fig:histograms}, and is centered at a velocity of 59 km s$^{-1}$, while the second structure has larger ($\sim 3$ km s$^{-1}$) line widths and shows a velocity gradient from 56 km s$^{-1}$ in the vicinity of the F1 core to over 58 km s$^{-1}$ in the vicinity of the F2 core. Whereas in IRDC C, the C1-N and C1-S velocity difference appears to be due to the cores residing in separate gas structures, the F1 and F2 cores (at 56.1 km s$^{-1}$ and 57.7 km s$^{-1}$, \citealt{Tan13}) appear to both lie in the same gas structure, with their velocity difference only being due to a velocity gradient in their parent gas structure. The higher J CO lines detected from Herschel all seem to correspond to this gas structure containing the F1 and F2 cores, with little high J CO emission being associated with the 59 km s$^{-1}$ gas structure towards the F1 clump. This correspondence of the higher J CO emission with one particular velocity structure, rather than occurring at intermediate velocities, suggests that this higher J CO emission is not formed from the interaction of the two velocity components, at least towards the F1 clump. 
	
	The detection of multiple velocity components within both low mass (e.g., \citealt{Hacar13, Pon14Kaufman}) and high mass star-forming regions (e.g., \citealt{Henshaw13, Henshaw14}) seems to be very common, such that it is unsurprising that IRDCs C and F show multiple CO components. In particular, IRDC F is confirmed to contain at least three separate velocity components, as traced by the optically thin C$^{18}$O J = 1 $\rightarrow$ 0 and 2 $\rightarrow$ 1 transitions, as well as by the N$_2$H$^+$ J = 1 $\rightarrow$ 0 transition (Barnes et al.~in preparation). \citet{Foster14} have also previously noted that the region around the F1 clump exhibits two separate velocity components. The detection of these multiple line components towards IRDC F, as well as the agreement of the N$_2$D$^+$ centroid velocity of the F1 core with one of our detected velocity components strongly suggests that the multiple components we detect in IRDC F are not just formed from self-absorption of the $^{13}$CO J  = 3 $\rightarrow$ 2 line, but rather trace different gas structures.
	
In both IRDC C and F, most of the detected $^{13}$CO components have FWHM of the order of 3 to 4 km s$^{-1}$. For an optical depth of the order of 3, opacity broadening will increase the observed line width by a factor of 1.5 \citep{Phillips79}. While this would decrease the FWHM of most detected components in the primary peak in the FWHM histograms to approximately 2 km s$^{-1}$, these lines must still contain a supersonic, non-thermal component, given that the expected thermal FWHM at 15 K is 0.15 km s$^{-1}$ for $^{13}$CO and the sound speed is 0.2 km s$^{-1}$. 

While some of the narrow detected components are reasonably weak, such that the narrow observed line widths could just be an artifact of a low signal to noise detection, many of these thin lines have reasonably large peak intensities, particularly for the 59 km s$^{-1}$ gas towards the F1 clump. We consider the narrow line widths of these gas components to be real, suggesting that these gas structures are dynamically less turbulent. While dense cores exhibit a transition to coherence \citep{Pineda10} that leads to reduced line widths as the cores evolve, this is unlikely to be the only cause of these small line widths as the C1-S and F1 cores are embedded within the gas structures with the larger FWHM. Small line widths can also be obtained at stagnation points, where two converging turbulent flows meet, since these stagnation points are where the relative velocities are minimum (e.g., \citealt{Klessen05}). However, stagnation points are where the compression of the gas is at a maximum, and thus is where cores should be located, which is again contrary to the F1 core being associated with the gas with larger line widths. Such narrow line widths could be also explained from the natural evolution of a dense core inside a globally infalling cloud \citep{NaranjoRomero15}.

The two velocities detected towards the F1 side of the map appear to form a bubble shape in Fig.~\ref{fig:f13coppv}. This bubble may have been shaped by the low mass protostars associated with the 24 micron source just to the north of the F1 core (see Fig.~\ref{fig:sigmaradec}), and the F1 core may have formed due to the compression of gas in the bubble wall, since the F1 core has a velocity consistent with the lower velocity edge of the bubble. Alternatively, this PPV distribution could be indicative of the merger of two gas components, rather than the formation of two components during the creation of a bubble structure. There is, however, no noticeable increase in FWHM at the location where these two components merge, as might be expected from the collision of two gas structures. Similarly, while the depletion of CO can cause a bifurcation in the centroid velocities of a gas structure \citep{Gomez14}, we do not think CO depletion is the cause of this bubble structure, as the N$_2$D$^+$ J = 3 $\rightarrow$ 2 line would otherwise have been expected to occur at an intermediate velocity, rather than at the same velocity as the lower velocity edge of the shell \citep{Tan13}. 

The $^{12}$CO and $^{13}$CO J = 3 $\rightarrow$ 2 lines can be used to trace outflows, with the emission from such outflows appearing as blue and redshifted line wings in spectra. An initial search for outflows in the JCMT data was performed by examining, by eye, successive channel maps of the data using the GAIA software package in order to search for obvious linear features in the data. Due to the complex morphology of the observed regions and insufficient angular resolution, no outflows were identified this way. Furthermore, no obvious outflow features were identified based upon the spectral line shapes, although such an identification would have been extremely difficulty given the alterations of the line profiles due to CO depletion, self-absorption, noise, and the presence of multiple velocity components.

\section{CONCLUSIONS}
\label{conclusions}

Four square arcminute maps of the $^{12}$CO, $^{13}$CO, and C$^{18}$O $\mbox{J} = 3 \rightarrow 2$ transitions were made towards two select locations within IRDCs C and F (G028.37+00.07 and G034.43+00.24) using the James Clerk Maxwell Telescope. These two maps contain the C1, F1, and F2 clumps, which contain the C1-N, C1-S, F1, and F2 quiescent, starless cores. Single pointing observations of the  $^{13}$CO and C$^{18}$O J = 2 $\rightarrow$ 1 transitions were also obtained towards the F1, F2, and C1-N cores with the IRAM 30m telescope.

The excitation temperature of the $^{12}$CO J = 3 $\rightarrow$ 2 line is of the order of 10 to 20 K throughout the JCMT observed fields, with a mean temperature of 13.4 K in IRDC C and 14.6 in IRDC F. Modelling of the multiple detected $^{13}$CO, and C$^{18}$O lines towards the F1, F2 and C1-N cores, however, indicate that these less abundant isotopologues can trace gas with kinetic temperatures up to 4 K lower than that traced by the $^{12}$CO J = 3 $\rightarrow$ 2 line. Similarly, the excitation temperature of the $^{13}$CO J = 3 $\rightarrow$ 2 line can be a few Kelvin less than the kinetic temperature of the $^{13}$CO emitting gas within IRDCs. 

Throughout IRDC C and F, significant levels of CO depletion are observed. The peak depletion towards IRDC F is within the range of 5 to 9, consistent with levels towards low mass star-forming cores \citep{Caselli99, Crapsi05}. The CO depletion in IRDC C is overall larger, with a maximum CO depletion value lying between 16 and 31. The level of depletion seen in IRDC C, however, is still consistent with large depletion factors found in other IRDCs \citep{Fontani12}.

The $^{12}$CO J = 3 $\rightarrow$ 2 spectra are highly optically thick and show significant self-absorption features. The IRDC C spectra typically have larger blue peaks, while the IRDC F spectra have larger red peaks. Such asymmetry is usually ascribed to large-scale infall and outflow motions, respectively, but such a simple interpretation may not hold for such large, complex regions as these IRDCs. 

The observed IRDCs show complex kinematical structure, with numerous velocity components detected towards multiple lines of sight. The $^{13}$CO J = 3 $\rightarrow$ 2 data reveal that the C1-N and C1-S cores are associated with two different gas structures, separated by approximately 2 km s$^{-1}$ in their velocity centroids. The higher velocity component, associated with C1-N, exhibits narrow, 1 km s$^{-1}$ line widths while the lower velocity component, associated with C1-S, has larger, 3 km s$^{-1}$ line widths. Towards IRDC F, we find that the F1 and F2 cores are embedded within the same velocity component, but one in which a significant velocity gradient exists. We detect a second velocity component towards the F1 clump, which creates a shell like structure in position-position-velocity space. This second component, unassociated with the F1 or F2 cores, tends to be thin, with FWHM of approximately 1 km s$^{-1}$. 

Overall, we find that IRDCs appear to be dynamically evolving structures with complex internal kinematics that yield dense cores with considerable depletion factors.

\begin{acknowledgements}
	We would like to thank our anonymous referee for helping improve the clarity and quality of this paper. The authors would like to thank Dr.~N.~Bailey, Dr.~J.~Bailey, Dr.~S.~Sadavoy, Dr.~J.~D.~Henshaw, and A.~Barnes for many insightful conversations regarding the data presented in this paper. The authors would also like to thank the Joint Astronomy Center for their assistance in reducing the JCMT data. Andy Pon and PC acknowledge the financial support of the European Research Council (ERC; project PALs 320620). DJ acknowledges support from a Natural Sciences and Engineering Research Council (NSERC) Discovery Grant. Aina Palau acknowledges financial support from UNAM-DGAPA-PAPIIT IA102815 grant, M\'exico. I.\ J.-S.\ acknowledges the financial support received from the STFC through an Ernest Rutherford Fellowship (proposal number ST/L004801/1). This research has made use of the Smithsonian Astrophysical Observatory (SAO) / National Aeronautics and Space Administration's (NASA's) Astrophysics Data System (ADS). This research has made use of the astro-ph archive. Some spectral line data were taken from the Spectral Line Atlas of Interstellar Molecules (SLAIM) (Available at http://www.splatalogue.net). (F. J. Lovas, private communication, \citealt{Remijan07}). More information regarding the GILDAS CLASS software can be found at http://www.iram.fr/IRAMFR/GILDAS. The James Clerk Maxwell Telescope has historically been operated by the Joint Astronomy Centre on behalf of the Science and Technology Facilities Council of the United Kingdom, the National Research Council of Canada and the Netherlands Organisation for Scientific Research.
\end{acknowledgements}

\bibliographystyle{aa}
\bibliography{ponbib}

\begin{appendix}

\section{JCMT SPECTRA}
\label{appendix:jcmtfits}

\begin{figure*}
   \centering
   \includegraphics[width=6.5in]{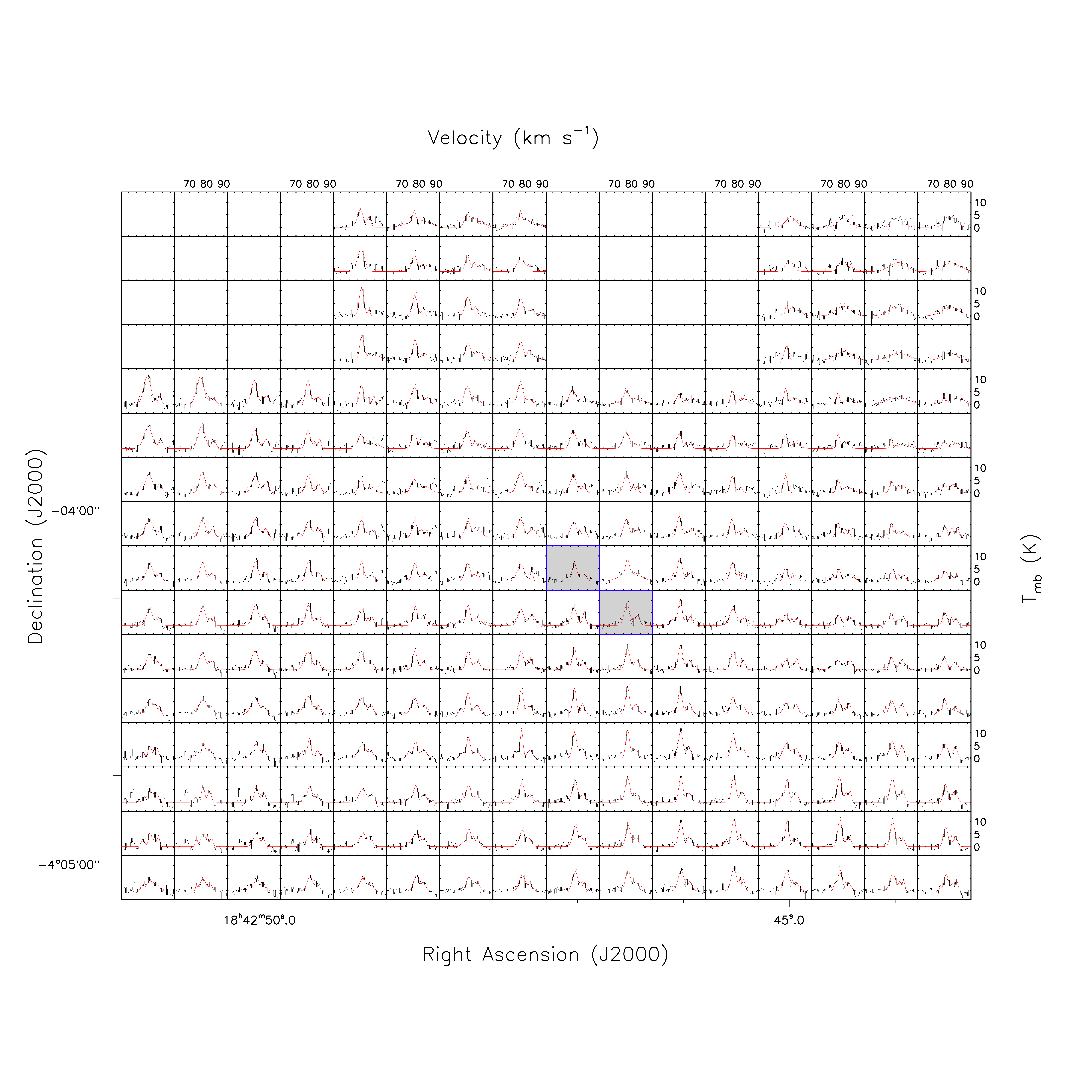}
   \caption{$^{12}$CO J = 3 $\rightarrow$ 2 spectra towards the C1 clump, as observed by the JCMT. The red lines show the cumulative fit of all detected components. The spectra outlined in blue and shown with grey backgrounds indicate the pixels closest to the C1-N and C1-S cores, with the C1-N core being the northern source.}
   \label{fig:c12coclinplotwfits}
\end{figure*}

\begin{figure*}
   \centering
   \includegraphics[width=6.5in]{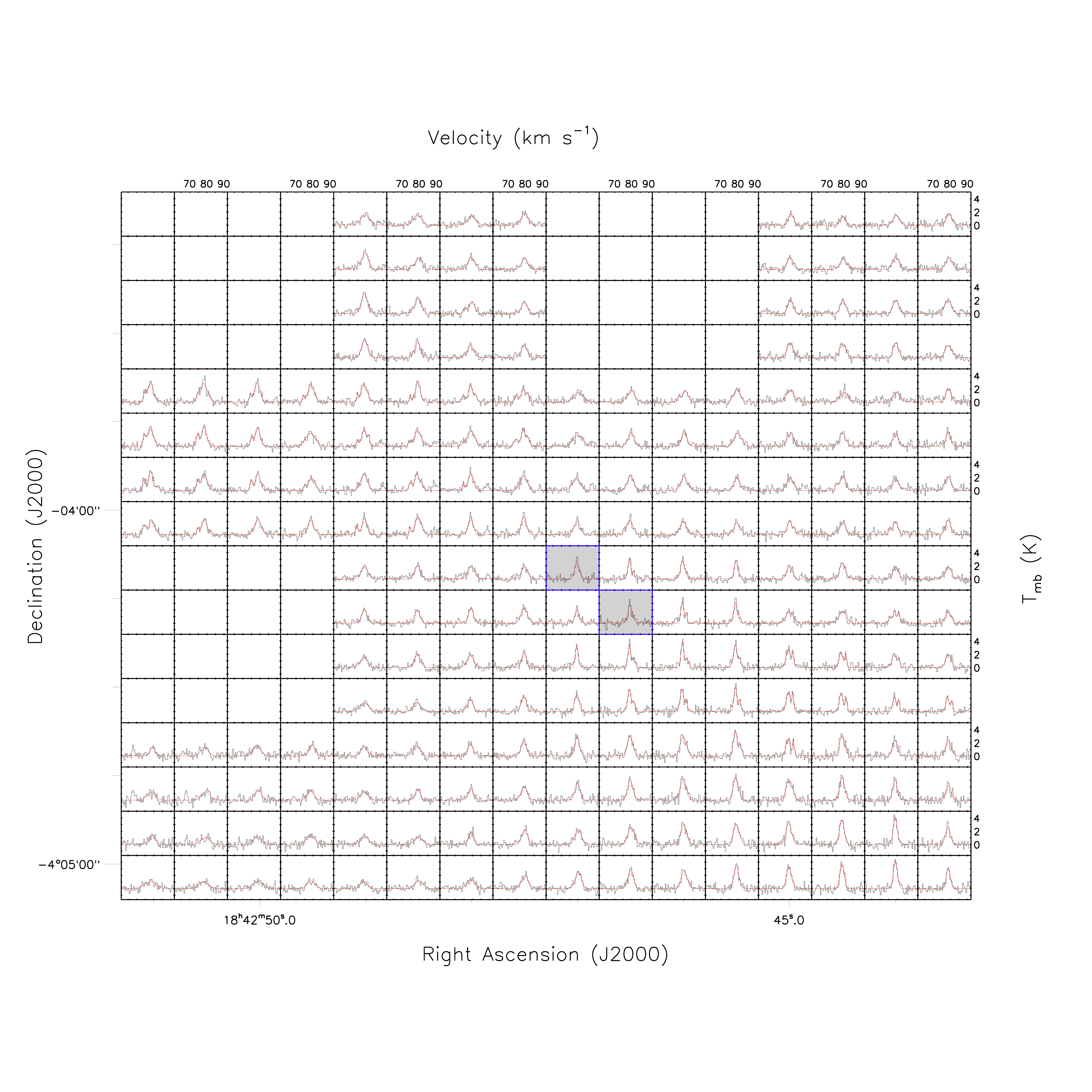}
   \caption{$^{13}$CO J = 3 $\rightarrow$ 2 spectra towards the C1 clump, as observed by the JCMT. The red lines show the cumulative fit of all detected components. The spectra outlined in blue and shown with grey backgrounds indicate the pixels closest to the C1-N and C1-S cores, with the C1-N core being the northern source.}
   \label{fig:c13coclinplotwfits}
\end{figure*}

\begin{figure*}
   \centering
   \includegraphics[width=6.5in]{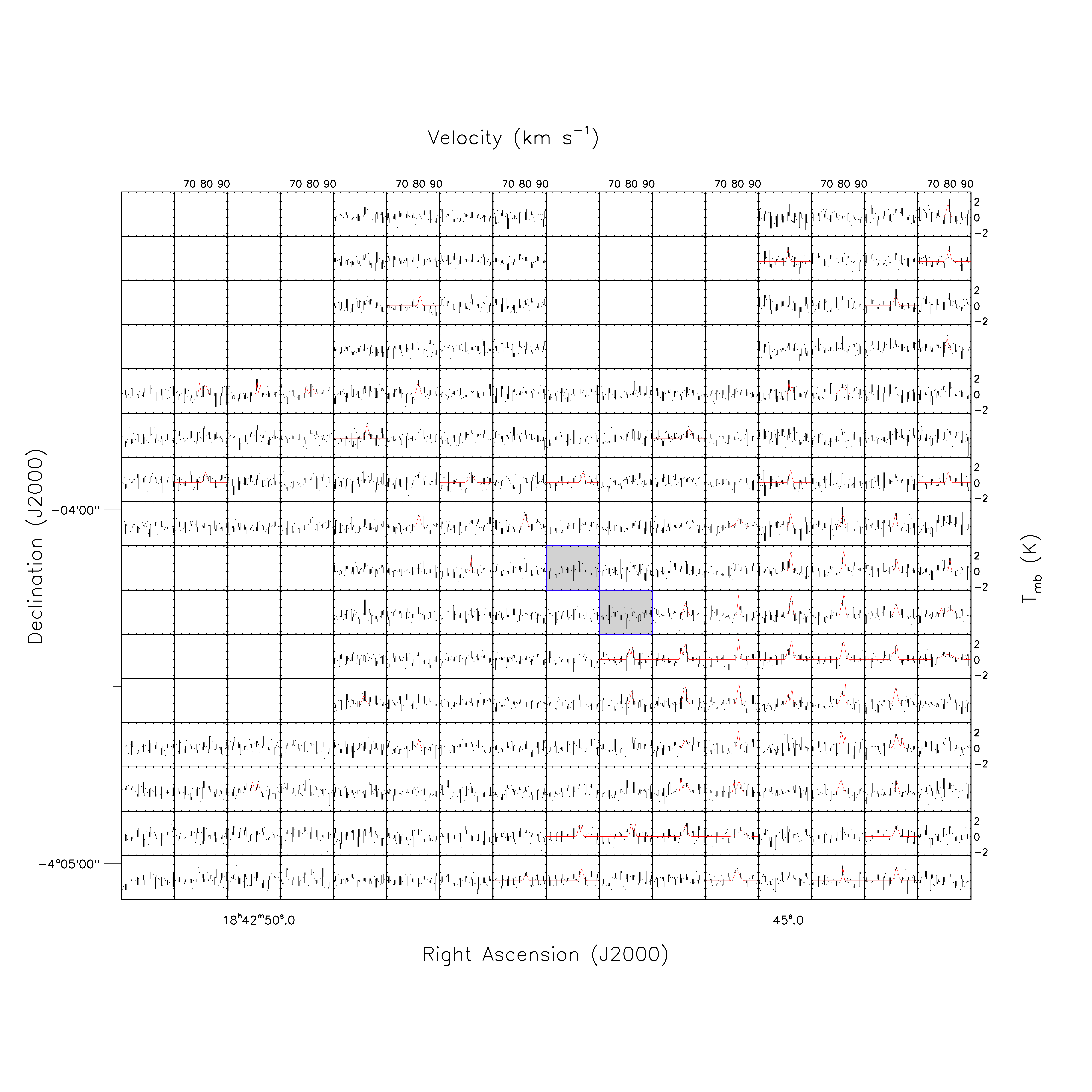}
   \caption{C$^{18}$O J = 3 $\rightarrow$ 2 spectra towards the C1 clump, as observed by the JCMT. The red lines show the cumulative fit of all detected components. The spectra outlined in blue and shown with grey backgrounds indicate the pixels closest to the C1-N and C1-S cores, with the C1-N core being the northern source.}
   \label{fig:cc18oclinplotwfits}
\end{figure*}

\begin{figure*}
   \centering
   \includegraphics[width=6.5in]{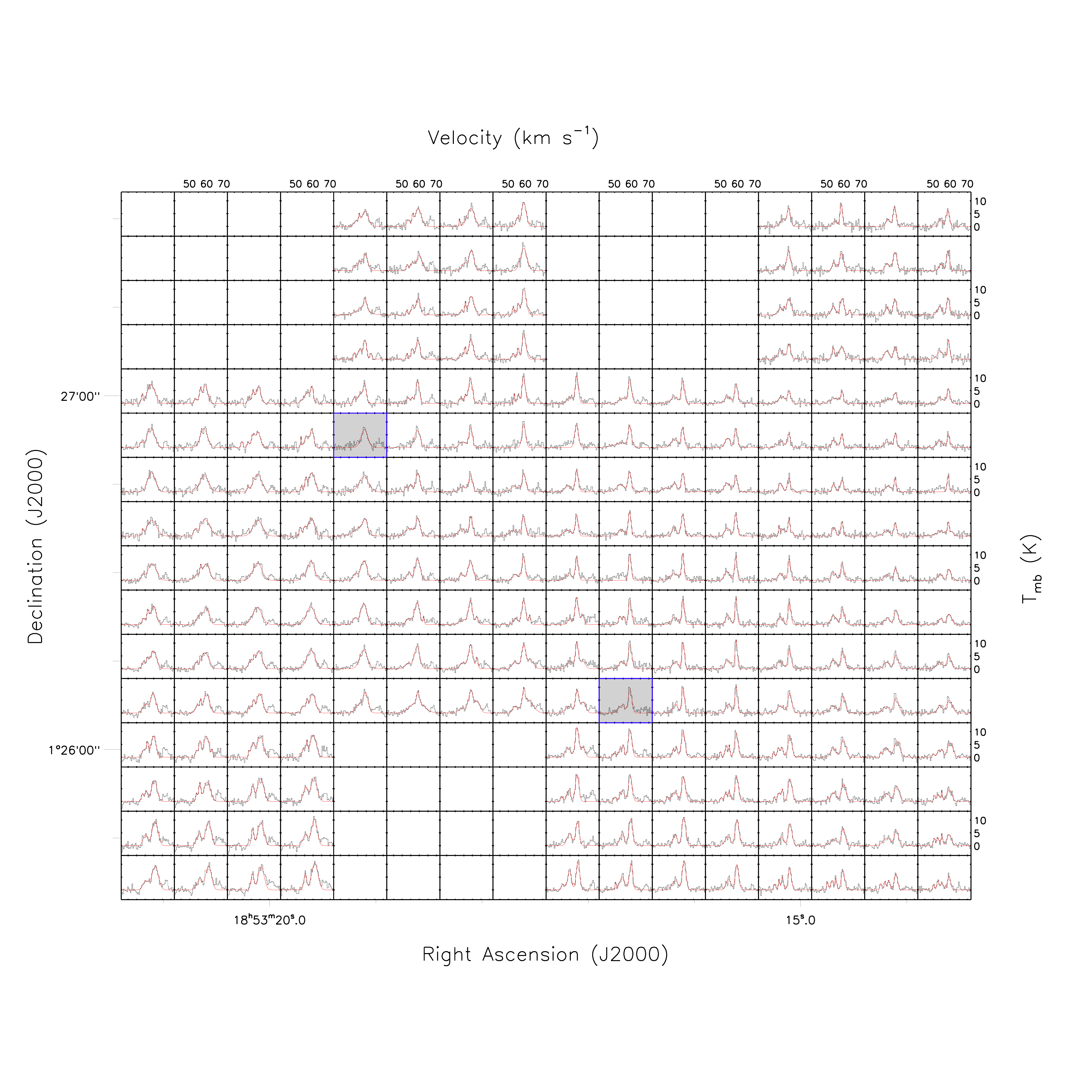}
   \caption{$^{12}$CO J = 3 $\rightarrow$ 2 spectra towards the F1 and F2 clumps, as observed by the JCMT. The red lines show the cumulative fit of all detected components. The spectra outlined in blue and shown with grey backgrounds indicate the pixels closest to the F1 and F2 cores, with the F2 core being the northern source.}
   \label{fig:f12coclinplotwfits}
\end{figure*}

\begin{figure*}
   \centering
   \includegraphics[width=6.5in]{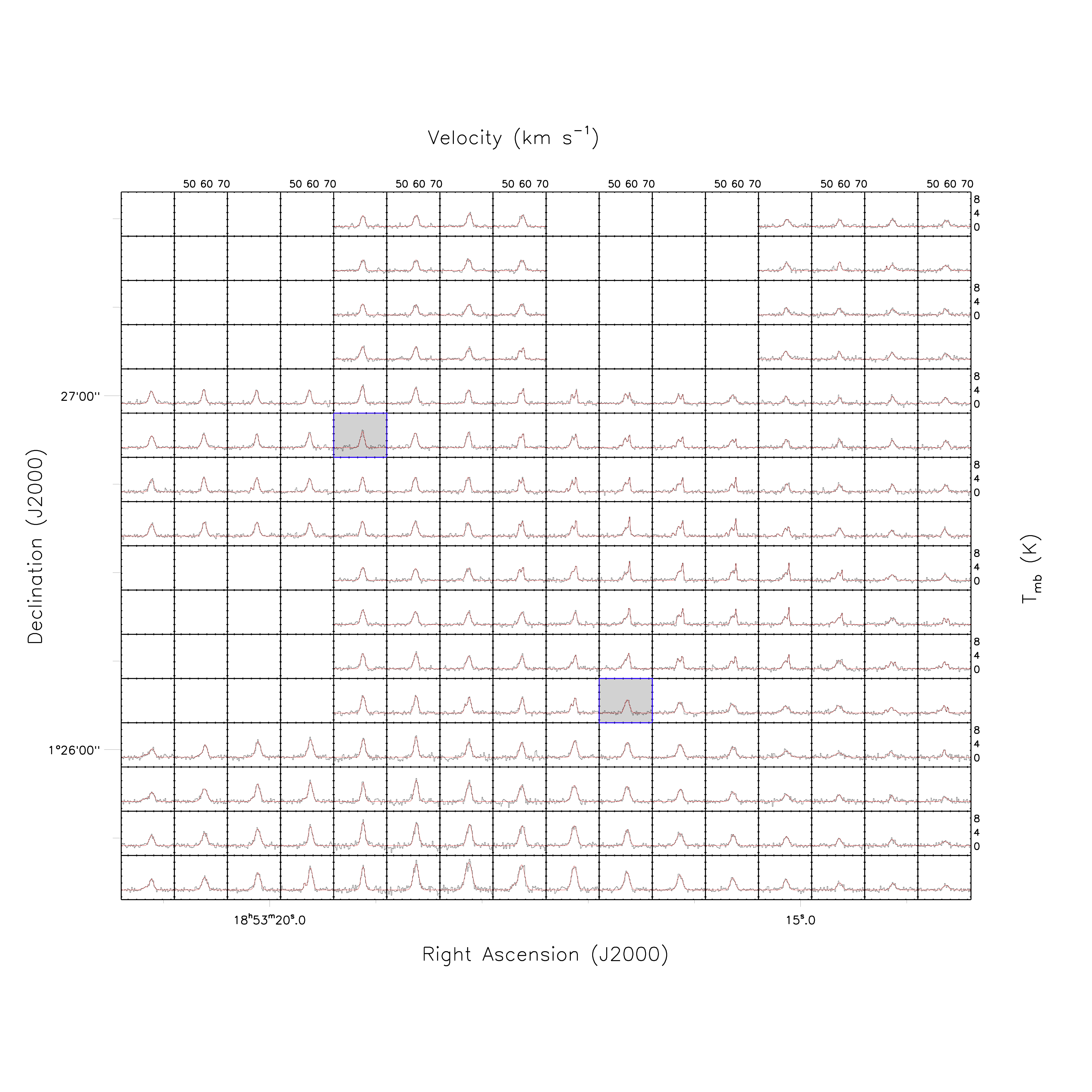}
   \caption{$^{13}$CO J = 3 $\rightarrow$ 2 spectra towards the F1 and F2 clumps, as observed by the JCMT. The red lines show the cumulative fit of all detected components. The spectra outlined in blue and shown with grey backgrounds indicate the pixels closest to the F1 and F2 cores, with the F2 core being the northern source.}
   \label{fig:f13coclinplotwfits}
\end{figure*}

\begin{figure*}
   \centering
   \includegraphics[width=6.5in]{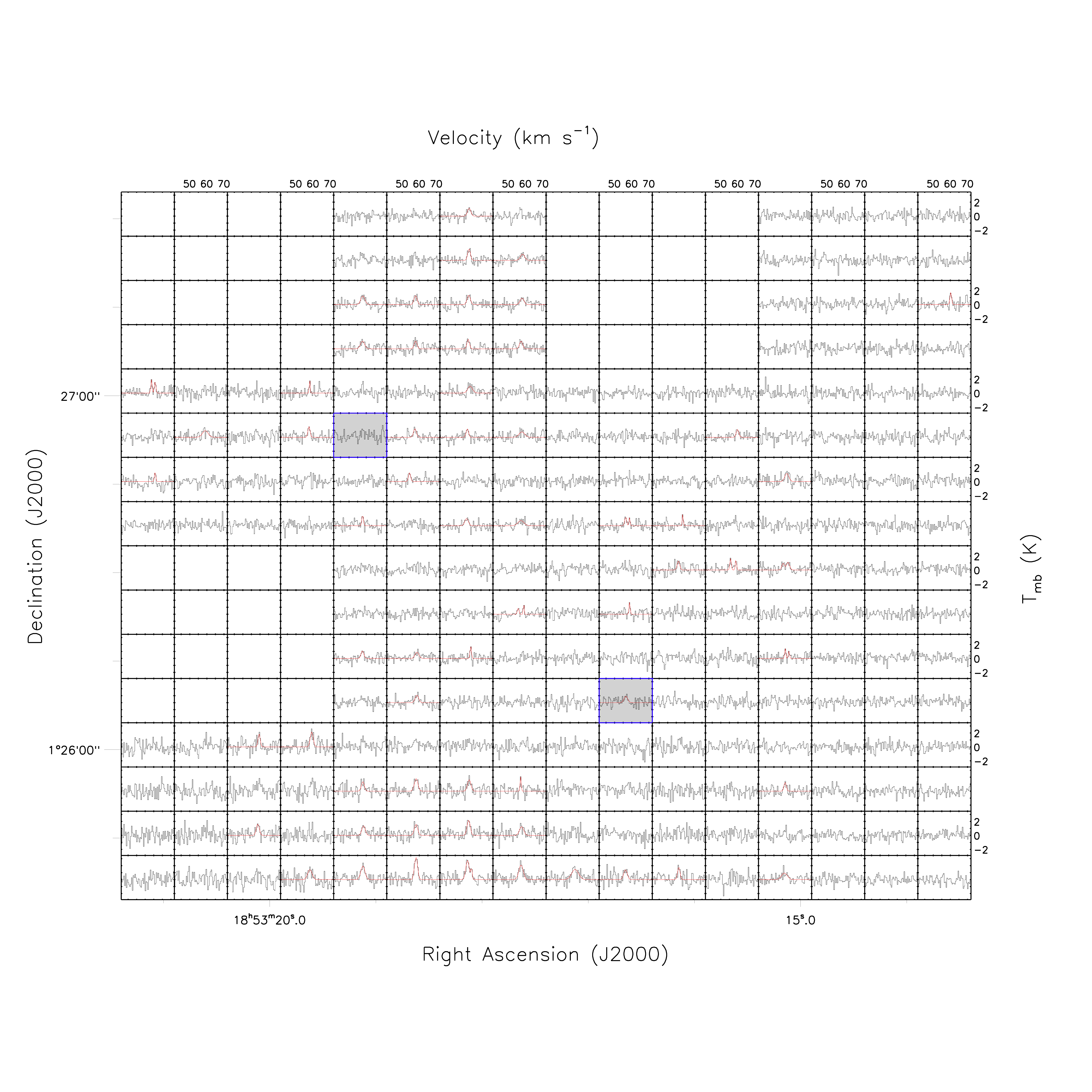}
   \caption{C$^{18}$O J = 3 $\rightarrow$ 2 spectra towards the F1 and F2 clumps, as observed by the JCMT. The red lines show the cumulative fit of all detected components. The spectra outlined in blue and shown with grey backgrounds indicate the pixels closest to the F1 and F2 cores, with the F2 core being the northern source.}
   \label{fig:fc18oclinplotwfits}
\end{figure*}

Figures \ref{fig:c12coclinplotwfits} to \ref{fig:fc18oclinplotwfits} show all of the spectra observed with the JCMT smoothed to a 0.5 km s$^{-1}$ velocity resolution. The red lines in the figures show the cumulative fit of all detected components.

\end{appendix}

\end{document}